\documentclass[a4paper,11pt]{article}
\pdfoutput=1 

\usepackage{jcappub} 
\usepackage[T1]{fontenc} 
\usepackage{bm}
\let\vec\bm

\newcommand{\T}{\mathbf{\hat{\mathcal{T}}}}

\newcommand{\Dk}[1]{\frac{d^3#1}{(2\pi)^3}}

\newcommand{\ve}[1]{{\text{\bf #1}}} 
\newcommand{\vk}{\vec k}
\newcommand{\vp}{\vec p}
\newcommand{\vq}{\vec q}
\newcommand{\vx}{\vec x}
\newcommand{\vv}{\vec v}
\newcommand{\vs}{\vec s}
\newcommand{\vu}{\vec u}
\newcommand{\vhn}{\hat{\vec n}}

\newcommand{\mA}{\mathcal{A}}
\newcommand{\mB}{\mathcal{B}}

\newcommand{\tm}{\text{m}}

\newcommand{\ikk}{\underset{\vk_{12}= \vk}{\int}}
\newcommand{\ikkk}{\underset{\vk_{123}= \vk}{\int}}

\newcommand{\dD}{\delta_\text{D}}
\newcommand{\Ps}{\mathbf{\Psi}}

\newcommand{\hmpc}{\,h^{-1}\text{Mpc}}
\newcommand{\hmpci}{\,h \, \text{Mpc}^{-1}}

\usepackage[dvipsnames]{xcolor}

\usepackage{booktabs}
\usepackage{float} 
\newcommand{\ra}[1]{\renewcommand{\arraystretch}{#1}}

\title{fkPT: Constraining scale-dependent modified gravity with the full-shape galaxy power spectrum}

\author[a]{Mario A.~Rodriguez-Meza,}

\author[b, 1]{Alejandro Aviles,\note{Corresponding author.}}
\emailAdd{aviles@icf.unam.mx}

\author[b,c,d]{Hernan E. Noriega,} 

\author[e]{Cheng-Zong Ruan,}

\author[f]{Baojiu Li,}

\author[c]{Mariana Vargas-Maga\~na,}

\author[a]{Jorge L.~Cervantes-Cota}

\affiliation[a]{Departamento de F\'isica, Instituto Nacional de Investigaciones Nucleares,
Apartado Postal 18-1027, Col. Escand\'on, Ciudad de M\'exico,11801, M\'exico.}

\affiliation[b]{Instituto de Ciencias F\'isicas, Universidad Nacional
Autónoma de México,  62210, Cuernavaca, Morelos.}

\affiliation[c]{Instituto de F\'isica, Universidad Nacional Autónoma de M\'exico, Apdo. Postal 20-364, 01000, D.F, México.}

\affiliation[d]{Institut de Ciències del Cosmos (ICCUB), Universitat de Barcelona, Martí i Franquès, 1, E08028 Barcelona, Spain.}

\affiliation[e]{Institute of Theoretical Astrophysics, University of Oslo, 0315 Oslo, Norway}

\affiliation[f]{Institute for Computational Cosmology, Department of Physics, Durham University, South Road, Durham DH1 3LE, UK}


\abstract{ Modified gravity models with scale-dependent linear growth typically exhibit an enhancement in the power spectrum beyond a certain scale. The conventional methods for extracting cosmological information usually involve inferring modified gravity effects via Redshift Space Distortions (RSD), particularly through the time evolution of $f\sigma_8$. However, classical galaxy RSD clustering analyses encounter difficulties in accurately capturing the spectrum's enhanced power, which is better obtained from the broad-band power spectrum. In this sense, full-shape analyses aim to consider survey data using comprehensive and precise models of the whole power spectrum. Yet, a major challenge in this approach is the slow computation of non-linear loop integrals for scale-dependent modified gravity, precluding the estimation of cosmological parameters using Markov Chain Monte Carlo methods. Based on recent studies, in this work we develop a perturbation theory tailored for Modified Gravity, or analogous scenarios introducing additional scales, such as in the presence of massive neutrinos. Our approach only needs the calculation of the scale-dependent growth rate $f(k,t)$ and the limit of the perturbative kernels at large scales. We called this approximate technique as fk-Perturbation Theory and implemented it into the code \texttt{fkpt}, capable of computing the redshift space galaxy power spectrum in a fraction of a second.  We validate our modeling and code with the $f(R)$ theory \textsc{MG-GLAM} and General Relativity \textsc{NSeries} sets of simulations. The code is available at \href{https://github.com/alejandroaviles/fkpt}{https://github.com/alejandroaviles/fkpt}.       }

\begin{document} 
\maketitle
\flushbottom

\begin{section}{Introduction}

During the last two decades galaxy surveys such as 2dF \cite{Cole:2005sx}, WiggleZ \cite{Blake:2011en}, BOSS \cite{Samushia:2013yga}, eBOSS \cite{Zhai:2016gyu}, and DES \cite{Abbott:2017wau} have contributed to the inference of cosmological parameters and the test of models using techniques complementary to other probes, including the Cosmic Microwave Background (CMB), Supernovae type Ia, among others. These analyses are expected to be even more important in the near future with the upcoming stage IV experiments, such as the \textit{Dark Energy Spectroscopic Instrument} (DESI) \cite{Aghamousa:2016zmz}, \textit{Euclid} \cite{EUCLID:2011zbd} and the \textit{Legacy Survey of Space of Time} (LSST) of the Vera C. Rubin Observatory \cite{Abate:2012za}, which are foreseen to achieve sub-percentage accuracy in various cosmological parameters. Therefore, models for the analysis are demanded to be sufficiently precise.

The traditional approach for analyzing galaxy clustering has centered on using a fixed template linear power spectrum and 
compress the data into a few parameters, $\alpha_\parallel$, $\alpha_\perp$ and $f \sigma_8$, accounting for the Alcock-Paczyński effect \cite{Alcock:1979mp} and Redshift Space Distortions (RSD) \cite{BOSS:2016wmc}. This method extracts the cosmological information from the Baryon Acoustic Oscillation (BAO) position and from the relative amplitude of the power spectrum multipoles, being insensitive to the rest of its features. Recently, this classical analysis has been extended to include a \textit{shape} parameter, $m$, that attempts to account for the broad-band piece of the spectrum \cite{Brieden:2021edu}. The advantage of these methods is their model independence, and hence the inferred parameters $f \sigma_8$ and cosmological distances as $d_A$ and $d_H$ are in principle valid for any theory reasonably close to $\Lambda$CDM. On the contrary, methods directly fitting a theoretical power spectrum, akin to the procedures in CMB analyses, have been present for  a long time; e.g. \cite{Montesano:2010qc,Montesano:2011bp,Angulo:2015eqa}. These theoretical frameworks has been much developed in different aspects of the nonlinear Perturbation Theory (PT) \cite{Bernardeau:2001qr}, including Effective Field Theory (EFT) \cite{McDonald:2006mx,McDonald:2009dh,Baumann:2010tm,Carrasco:2012cv,Vlah:2015sea}, nonlinear bias \cite{McDonald:2006mx,McDonald:2009dh,Assassi:2014fva,Senatore:2014eva,Lewandowski:2014rca,Desjacques:2016bnm}, and Infrared resummations (IR-resummations) \cite{Senatore:2014via,Baldauf:2015xfa,Lewandowski:2018ywf,Ivanov:2018gjr}, 
paving the way for modern analysis methods. In this regard, the implementation of a full-shape, also called full-modeling or direct-fit, analysis of the power spectrum has emerged as a primary tool to determine the cosmological model from data, particularly since the works \cite{Ivanov:2019pdj,DAmico:2019fhj}. Subsequent research has expanded this methodology, both within the standard vanilla $\Lambda$CDM \cite{Nishimichi:2020tvu,Philcox_2020,Chen:2021wdi,Zhang:2021yna,Donald-McCann:2023kpx,Ramirez:2023ads} and for extended models, including the consideration of massive neutrinos \cite{Aviles:2021que,Noriega:2022nhf,Kumar:2022vee,Moretti:2023drg}, curvature \cite{Chudaykin:2020ghx,Glanville:2022xes}, or more exotic models as dark energy and Modified Gravity (MG) \cite{DAmico:2020kxu,DAmico:2020tty,Piga:2022mge,Moretti:2023drg,Euclid:2023bgs,Gsponer:2023wpm}, among others, e.g.,  \cite{He:2023oke,Camarena:2023cku}. Other theoretical aspects, especially  the use of different sets of parameters and priors have gained much attention recently \cite{Carrilho:2022mon,Simon:2022lde,Donald-McCann:2023kpx,Holm:2023laa}. Finally, a few codes have been released to compute the EFT-PT theoretical power spectrum in redshift space: \texttt{Class-pt} \cite{Chudaykin:2020aoj},\footnote{\href{https://github.com/Michalychforever/CLASS-PT}{https://github.com/Michalychforever/CLASS-PT}} \texttt{Velocileptors}  \cite{Chen:2020fxs,Chen:2020zjt},\footnote{\href{https://github.com/sfschen/velocileptors}{https://github.com/sfschen/velocileptors}} \texttt{PyBird} \cite{DAmico:2020kxu},\footnote{\href{https://github.com/pierrexyz/pybird}{https://github.com/pierrexyz/pybird}} and \texttt{FOLPS-nu} \cite{Noriega:2022nhf}.\footnote{\href{https://github.com/henoriega/FOLPS-nu}{https://github.com/henoriega/FOLPS-nu}} These codes use FFTLog methods \cite{1978JCoPh..29...35T,Hamilton:1999uv,McEwen:2016fjn,Fang:2016wcf,Schmittfull:2016jsw,Schmittfull:2016yqx,Simonovic:2017mhp} to accelerate the computation of loop integrals.

The above-mentioned surveys can be employed to test gravity. As it is known, General Relativity (GR) is well tested in local, planetary scales using for instance Parametrized Post Newtonian (PPN) parameters \cite{Will:2014kxa}, and several other approaches in astrophysical scales \cite{Baker:2019gxo}. This seems to be the case also in Cosmology, since the standard $\Lambda$CDM model is in general consistent with current cosmological data, and  GR is believed to be correct at large scales. 
Most of the gravity theory tests so far have employed RSD, since this effect is intimately related to the growth rate of matter perturbations that depends upon the gravity model. RSD has been developed for many years, from the pioneering work of Kaiser \cite{10.1093/mnras/227.1.1} to modern nonlinear perturbation theory, see e.g. \cite{Vlah:2018ygt,Chen:2020zjt,Lewandowski:2015ziq,Perko:2016puo}. However, with the advent of more precise Large Scale Structure (LSS) data, new opportunities to test gravity at cosmological scales arise, yet the analysis demands methods of non-linear PT, such as the full-shape technique. This is because one of the main potential observables of MG theories is an enhancement of the power spectrum above a certain scale due to a larger strength of gravity, hence a major impact over the power spectrum is expected to lie in its broad-band shape. 
Let us shortly discuss about it. The sound horizon at decoupling is practically the same for $\Lambda$CDM and viable MG models, since they only affect late-time physics, and hence the BAO scale is unaltered, with the exception of a small extra degradation due to differences in the large scale displacement fields \cite{Aviles:2017aor}. On the other hand, RSD analysis measures the amplitude of fluctuations $f \sigma_8$ by breaking the degeneracy of the linear bias with the relative size between the different multipoles of the power spectrum. It is for this reason that the standard analysis (BAO + RSD) alone, we claim, is not efficient for detecting an MG signal, but one still needs the information carried by the broad-band of the power spectrum, that is precisely provided by the full-shape analysis.

The PT of MG has been explored extensively over the last decade \cite{Koyama:2009me,Taruya:2013quf,Brax:2013fna,Bellini:2015oua,Taruya:2014faa,Taruya:2016jdt,Aviles:2017aor,Winther:2017jof,Fasiello:2017bot,Bose:2016qun,Bose:2017dtl,Bose:2018orj,Valogiannis:2019xed,Valogiannis:2019nfz,Aviles:2018qotF,Aviles:2018saf}. Recently, a few of us \cite{Aviles:2020wme} proposed an accurate theoretical PT/EFT for MG models having a scale-dependent linear growth. This model was tested against the \textsc{Elephant} suite of N-body simulations for $\Lambda$CDM and the Hu-Sawicki (HS) $f(R)$ model \cite{Hu:2007nk}, finding very good agreement up to $k \sim 0.2 \, h\,\text{Mpc}^{-1}$ at redshift $z = 0.5$. Unfortunately, the computation of non-linear corrections to the power spectrum within this methodology is quite slow since the perturbative kernels are not known analytically, as for the $\Lambda$CDM case. Instead, they should be obtained from a set of differential equations that not only depend on the wave vectors configuration but also on the cosmological parameters. That is, to find the one-loop corrections to the power spectrum, having a form similar to
\begin{equation}
I(k) = \int d\vp \, \mathcal{K}(\vk,\vp) P_L(k) P_L(|\vk-\vp|),
\end{equation}
one must solve differential equations to find the kernels $\mathcal{K}$ at each volume element of the integration. This complexity hinders the use of a Monte Carlo Markov Chain (MCMC) analysis for parameter estimation. A similar challenge arises in cosmological scenarios involving massive neutrinos, where scale-dependence is introduced through free-streaming and has been addressed in \cite{Aviles:2021que,Noriega:2022nhf}. In this work we build upon the ideas of these references and develop a method that is able to obtain the power spectrum in a fraction of a second. To do this, we first identify three effects that modify the standard Einstein-de Sitter (EdS) 
kernels: 
\begin{itemize}
 \item[\textit{i)}] The failure of the velocity field $\theta = \nabla\cdot \vec v /(aHf)$, to be equal to the overdensity $\delta$ at the linear perturbation level because the growth rate $f$ at a given time is no longer constant, as in $\Lambda$CDM, but becomes scale-dependent; see fig.~\ref{fig:fk} below. The key relation is given by eq.~\eqref{theta1delta1}, which also enters into the non-linear kernels. 

 \item[\textit{ii)}] The abundance of matter $\Omega_m$ cannot be approximated by the square of the growth rate, $f^2$. This is particularly true for gravity models with scale-dependent linear growth, but may also be the case for some only time-dependent modified gravity models. 

 \item[\textit{iii)}] The presence of non-linear screenings that should drive the MG theories at small cosmological scales.
\end{itemize}

The effects \textit{ii)} and \textit{iii)} give rise to the differential equations that considerably slow the computation of the one-loop power spectrum. Fortunately, we realize that the breakdown of the approximation $\Omega_m \neq f^2$ is not very harmful and can be healed by taking only time-dependent corrections to the EdS kernels, as it is shown below in fig.~\ref{fig:ScreeningsVsEFT}.
Furthermore, the scale-dependent screenings operate on scales where the theoretical power spectrum is dominated by the EFT counterterms and become widely degenerate with them. We show that this is the case for HS-$f(R)$ gravity model,\footnote{See fig.~\ref{fig:ScreeningsVsEFT} in \S \ref{sec:fkPerturbationTheory} for a comparison between EFT contributions and screenings.} but as long the screenings are small at mildly non-linear scales and below, this can be a good approximation for any MG theory. Although the validity of this assumption is expected to break above some wave-number, and should be preferably checked case by case. Finally, in \S \ref{sec:fkPerturbationTheory} we notice that the dominant contribution is due to the effect described in point \textit{i)}. This allows us to construct a method that first evaluates $f(k,t)$ using MG linear theory and then the perturbative kernels are evaluated ignoring the effects of \textit{ii)} and \textit{iii)}. This method was proposed initially for massive neutrinos in \cite{Aviles:2021que} and further developed in \cite{Noriega:2022nhf} where the Python code \texttt{Folps-nu} was released. In the present work we test this methodology for the HS-$f(R)$ theory, particularly we validate it against the state-of-the-art \textsc{MG-GLAM} simulations \cite{Hernandez-Aguayo:2021kuh,Ruan:2021wup}. Once we ensure that our method is able to detect the MG signal, we test it against high precision \textsc{NSeries} simulations, which exists only for $\Lambda$CDM and were widely used in the past to test the pipeline of BOSS survey \cite{BOSS:2016wmc}.

Together with this paper we release the \textsc{C} language code \texttt{fkpt},\footnote{\href{https://github.com/alejandroaviles/fkpt}{https://github.com/alejandroaviles/fkpt}} that computes the redshift space one-loop power spectrum of MG theories in a fraction of a second, including biasing terms, EFT counterterms, shot noise, as well as IR-resummations. Contrary to the PT codes enlisted above, \texttt{fkpt} does not use FFTLog methods since we want a flexible method that allows for a future incorporation of theories that do not reduce to $\Lambda$CDM at very large scales, and hence their kernels cannot be approximated as EdS when they are evaluated at small wave vectors. This, for example, is done for the scale-independent nDGP gravity model in \cite{Piga:2022mge}. As explained in \cite{Aviles:2021que}, the large scale kernels modifications almost double the number of matrix multiplications that should be done if using an FFTLog method. Hence, we adopt here a brute force approach, but equally accurate and almost as fast as FFTLog based codes, for solving the loop integrals. We expect our perturbative method and our code can be useful for testing gravity with the full-shape galaxy power spectrum of future spectroscopic surveys, particularly for DESI and Euclid. 

The rest of this work is organized as follows: In \S \ref{sect:MG_model} we discuss the basic equations of the $f(R)$ model that are relevant to us. In \S \ref{sect:theory} we delve into the general non-linear theory for models with scale-dependent linear growth, putting special emphasis on MG. In \S \ref{sec:fkPerturbationTheory} we present the relevant approximations that lead to the fk-Perturbation Theory as well as our code \texttt{fkpt}. In \S \ref{sect:model_validation} we validate our methodology using the \textsc{MG-GLAM} and \textsc{NSeries} simulations. Some of the formulas are displayed in a separate appendix. Final remarks and conclusions are put forward in \S \ref{sect:conclu}.

\end{section}

\begin{section}{Modified gravity} \label{sect:MG_model}

Although the perturbation theory method presented throughout this work is very general and encompasses a large variety of MG theories, we will concentrate on the case of $f(R)$ gravity \cite{Carroll:2003wy,Hu:2007nk}, which is perhaps the most studied scale-dependent MG model in Cosmology and for which we have the most precise simulations. It is defined through the action
\begin{equation} \label{f_R_action}
S= \frac{1}{16\pi G}\int d^4 x \sqrt{-g} \left( R + f(R) \right) +  S_m(\psi^i,\ve g),
\end{equation}
where $R$ is the Ricci scalar and a function $f(R)$ is added to the Einstein-Hilbert GR action. Further, $S_m$ the action of matter fields $\psi^i$ and depends also on the metric $\ve g$. 
Variations with respect to the metric of this action lead to the field equations
\begin{equation}\label{fRFE}
 G_{\mu\nu} + f_R R_{\mu\nu} - \nabla_\mu \nabla_\nu f_R - \left( \frac{f}{2} - \square f_R \right)g_{\mu\nu} = 8 \pi G T_{\mu\nu},   
\end{equation}
where $f_R = \partial f(R) /\partial R$, and with energy momentum tensor
\begin{equation}
    T_{\mu\nu} = - \frac{2}{\sqrt{-g}} \frac{\delta S_m}{\delta g^{\mu\nu}}.
\end{equation}

We use the Friedmann-Robertson-Walker metric with additional scalar perturbations in Newtonian gauge 
\begin{equation}
ds^2 = - (1+2\Phi) dt^2 + a(t)^2 (1-2\Psi) d\vx^2, 
\end{equation}
considering the fluid perturbation $\Delta \rho = \bar{\rho} \delta$ and the MG associated scalar field perturbation  $ \delta f_{R} = f_R - \bar{f_R}$, $R= \bar{R} + \delta R$, where the bar indicates background quantities, and $\bar{R} \equiv R(\bar{f}_R)$. The perturbative field equations in Fourier space are \cite{Koyama:2009me} 
\begin{eqnarray}
- \frac{k^2}{a^2} \Phi &=& 4 \pi G \bar{\rho} \delta + \frac{1}{2} \frac{k^2}{a^2} \delta f_{R},
\label{PoissonEq_fR} \\
 \frac{k^2}{a^2} \delta f_{R} 
&=&  \frac{8 \pi G}{3} \bar{\rho} \delta - \frac{M_1(k)}{3} \delta f_{R}  - \frac{{\delta \cal I}(\delta f_{R})}{3}, 
\label{KG_fR}
\\
  \Psi - \Phi &=&  \delta f_{R}.  
\end{eqnarray}
Equation \eqref{PoissonEq_fR} is the Poisson equation, modified by the term $\delta f_{R}$, which then acts as a fifth force. We have defined 
\begin{equation}
  M_1 \equiv \frac{d R(f_R)}{d f_{R}} \Bigg|_{f_R = \bar{f}_R},  
\end{equation}
 such that eq.~\eqref{KG_fR} governs the evolution of the scalar field, a.k.a. the Klein-Gordon equation, from which we can read that the mass of scalar field $m = (M_1 /3)^{1/2}$, such that one recovers the $\Lambda$CDM model for scales larger than $m^{-1}$. 
 
Using eq.~(\ref{KG_fR}) to eliminate $\delta f_{R}$ from  eq.~(\ref{PoissonEq_fR}), we arrive at 
\begin{equation}\label{PoissonEq}
    -\frac{k^2}{a^2}\Phi = A(k,t) \delta(\vk) + \mathcal{S}(\vk),
\end{equation}
with
\begin{align}
A(k,t) &\equiv 4 \pi G \bar{\rho} \, \mu(k,t), \label{A_def} 
\end{align}
and
\begin{align}
\mu(k,t) &= 1 + \frac{1}{3} \frac{k^2}{k^2 +  a^2m^2(a)} , \label{mu_Eq} \\
\mathcal{S}(\vk) &= -\frac{1}{6}  \frac{k^2}{k^2 +  a^2m^2(a)}  {\delta \cal I}.
\label{S_Eq}
\end{align}
As $k \gg m$, the function $ \mu$ goes to $4/3$, and hence, in $f(R)$ theories the strength of gravity is enhanced by a factor of one third at the smallest scales.  This, of course may rule out any $f(R)$ theory. However the nonlinear interactions, collected in the term ${\delta \cal I}$ provide a screening mechanism responsible to drive the theory to GR at specific limits. The screening mechanism behind $f(R)$ and other scalar-tensor theories is the chameleon \cite{Khoury:2003aq,Khoury:2003rn}, that turns off the fifth force in regions with sufficiently deep potentials. In perturbation theory one expands the non-linear self-interaction $\delta \mathcal{I}$ as\footnote{We use the shorthand notation
\begin{equation}
  \underset{\vk_{1\cdots n}= \vk}{\int} = \int \Dk{k_1} \cdots  \Dk{k_n} (2\pi)^3\dD(\vk_{1\cdots n}-\vk)  
\end{equation}
and $\vk_{1\cdots n} = \vk_1 +\cdots + \vk_n$.
} 
\begin{align}\label{selfIntexp}
\delta \mathcal{I}(\delta f_{R})
&= \frac{1}{2} \ikk M_2(\vk_1, \vk_2)
\delta f_{R}(\vk_1) \delta f_{R}(\vk_2) \nonumber\\
&\quad + \frac{1}{6}
\ikkk M_3(\vk_1, \vk_2, \vk_3)
\delta f_{R}(\vk_1) \delta f_{R}(\vk_2) \delta f_{R}(\vk_3) + \cdots,
\end{align}
where the functions $M_i$  are in general scale and time dependent. For the $f(R)$ theories, these are only time dependent and given by the coefficients of the Taylor expansion 
\begin{equation} \label{potexp}
 \delta R = \sum_i \frac{1}{n!} M_n (\delta f_R)^n, \qquad M_n \equiv \frac{d^n R(f_R)}{d f_{R}^n} \Bigg|_{f_R = \bar{f}_R}. 
\end{equation}

Throughout this work we apply our results to the specific case of HS-$f(R)$ model \cite{Hu:2007nk}, with $n=1$, defined by   
\begin{equation}\label{HSfR}
 f(R)= - \frac{c_1 R}{c_2 R/M^2 + 1}.
\end{equation}

In order to have an effective $\Lambda$CDM model at background level, the energy scale is chosen to be $M^2= \Omega_{m} H_0^2 $ and the remaining constants comply with $c_1/c_2 = 6 \Omega_\Lambda / \Omega_{m}$. 

The functions $M_1$, $M_2$ and $M_3$ in HS-$f(R)$ are
\begin{align}
M_1(a) &= \frac{3}{2}  \frac{H_0^2}{|f_{R0}|} \frac{(\Omega_{m} a^{-3} + 4 \Omega_\Lambda)^3}{(\Omega_{m}  + 4 \Omega_\Lambda)^2}, \label{M1} \\ 
M_2(a) &= \frac{9}{4}  \frac{H_0^2}{|f_{R0}|^2} \frac{(\Omega_{m} a^{-3} + 4 \Omega_\Lambda)^5}{(\Omega_{m}  + 4 \Omega_\Lambda)^4}, \label{M2} \\ 
M_3(a) &= \frac{45}{8}  \frac{H_0^2}{|f_{R0}|^3} \frac{(\Omega_{m} a^{-3} + 4 \Omega_\Lambda)^7}{(\Omega_{m}  + 4 \Omega_\Lambda)^6}. \label{M3}
\end{align}
These functions depend only on the background evolution since they are the coefficients of the expansion of a scalar field potential 
about its background value, see eq.~(\ref{potexp}).

The second order source entering eq.~\eqref{DAeveq} becomes 
\begin{equation}
\mathcal{S}^{(2)}(\vk_1,\vk_2) = -\left(\frac{8\pi G}{3} \right)^2 \frac{ M_2(\vk_1,\vk_2) k^2/a^2}{6 \Pi(k)\Pi(k_1)\Pi(k_2)},     
\end{equation} \label{defPi}
where $\Pi(k) \equiv \frac{1}{3 a^2}(3 k^2 + M_1 a^2)$.

    The strength of the MG is given by the amplitude $f_{R0}$, the smaller is the value $f_{R0}$, the smaller the effect of MG. As a matter of notation the models with $f_{R0} = -10^{-6},-10^{-5}$ are called F6 and F5, respectively. While $f_{R0}=0$ corresponds to GR.  

Finally, we stress that results for other gravity models are straightforward to develop following the methods of the present and previous works;
see e.g. Appendix B of \cite{Bose:2016qun}.
Moreover, theories defined in the Einstein frame can be put also within our framework by using field redefinitions \cite{Aviles:2018qotF}.

\end{section}

\begin{section}{Theoretical framework and perturvative kernels} \label{sect:theory}

In this section we present the general perturbative framework for theories with scale-dependent linear growth. To do this we first find the Lagrangian Perturbation Theory (LPT) kernels, afterwards we map them into the Eulerian frame, which is where we compute the multipoles of the power spectrum.

The LPT for MG was developed in \cite{Aviles:2017aor}. More recently, ref.~\cite{Aviles:2020cax} presented an LPT scheme for studying the clustering of dark matter particles in the presence of massive neutrinos. This framework is very general and can be applied for cosmologies with additional scales, and as such, under a few amendments it can be rewritten as a theory for MG, reducing to that of \cite{Aviles:2017aor}. In \S\ref{sec:LPT}-\S\ref{subsect:Lag_Disp} we present such scheme up to second order, since this is sufficient to understand the whole development idea, leaving the third order final results to Appendix \ref{app:3rdOrder}. After that, in \S \ref{subsect:LPTtoSPT} we show how to perform a mapping to obtain the kernels in Standard Perturbation Theory (SPT), that we use to obtain the power spectrum non-linear corrections. 
 
\begin{subsection}{Evolution in Lagrangian space} \label{sec:LPT}

In the Lagrangian framework there is a particular interest in the map between the initial (Lagrangian) positions $\vq$ of the cold dark matter particles and the final, or moment of observation, Eulerian positions $\vx$. This map between frames is given by the Lagrangian displacement vector field $\Ps$
\begin{equation}
     \vx(\vq,t) = \vq + \Ps(\vq,t). 
\end{equation}
The displacement evolves according to the Geodesic equation, 
\begin{equation} \label{geodesicEq}
   \T \, \Ps(\vq,t) \equiv \left(   \frac{d^2\,}{d t^2} +  2 H \frac{d\,}{d t} \right)\Ps(\vq,t)  = - \frac{1}{a^2} \vec{\nabla}_{\vx} \Phi(\vx,t) \Big|_{\vx=\vq+\Ps},
\end{equation}
where the first equality serves to  define the differential operator $ \T$ \cite{Matsubara:2015ipa}, and $\vec{\nabla}_{\vx} =\partial /\partial \vx $ are the derivatives with respect to the Eulerian coordinates. 
The gravitational potential $\Phi$ obeys the Poisson equation
\begin{equation}\label{Poisson}
 \frac{1}{a^2}\nabla^2_{\vx} \Phi(\vx,t) = 4 \pi G \rho_m \delta(\vx,t) + S(\vx,t), 
\end{equation}
which here is modified by the term $S(\vx,t)$. For example, in theories in the presence of massive neutrinos one identifies $\delta = f_{cb} \delta_{cb} $ and $S(\vx,t) = 4 \pi G \rho_m f_{\nu} \delta_{\nu}$ \cite{Aviles:2020cax}, where the label $cb$ refers to the combined baryons and cold dark matter fluid and the label $\nu$ to the neutrino component. Furthermore, the relative abundances are defined as $f_{cb}=\Omega_{cb}/\Omega_m $ and $f_{\nu}=\Omega_{\nu}/\Omega_m $. Also, in this case $\Psi \rightarrow f_{cb} \Psi$, so the Lagrangian displacement follows the $cb$ particles. 

In several MG theories one can write  $\delta = \delta_m $ and $S = \frac{1}{a^2}\nabla^2_{\vx} \phi$, where $\phi$ is an extra degree of freedom. We will follow this route below. 

Taking the divergence of eq.~\eqref{geodesicEq} and moving to Fourier space 
we write
\begin{equation}\label{eom}
 \Big[\nabla_{\vx} \cdot \T \Psi(\vq)\Big](\vk) = - 4 \pi G \rho_m \tilde{\delta}(\vk) - \tilde{S}(\vk), 
\end{equation}
where we omitted to write the time dependencies explicitly.
Here, $[(\cdots)](\vk)$ indicates the Fourier transform of $(\cdots)(\vq)$. That is, for a function $f(\vq)$ of the Lagrangian coordinates $\vq$, the notation means 
\begin{equation}
  \big[f(\vq) \big](\vk) = \int \Dk{k} e^{i\vk \cdot \vq} f(\vq).    
\end{equation}
Further, for notational convenience, we write a tilde over Fourier transforms as $ \tilde{\delta}(\vk)$ and  $ \tilde{S}(\vk)$ to point out that they are ``$q$-Fourier'' transforms of functions defined over the  Eulerian coordinates.  That is, for $f(\vx)$ we have
\begin{equation}\label{xFT}
 \tilde{f}(\vk) = \int d^3 q \, e^{i\vk \cdot \vq} f(\vx).    
\end{equation}
Finally, we note that the ``standard-Fourier''  transform  can be computed in any frame as 
\begin{align} \label{xFT2}
         f(\vk) &= \int d^3 x \, e^{-i\vk \cdot \vx} f(\vx) =  \int d^3 q \, e^{-i\vk \cdot \vq} f(\vq) = \big[f(\vq) \big](\vk).
\end{align}

Now, the equation of motion \eqref{eom} contains functions that take Eulerian coordinates as arguments and functions that take Lagrangian coordinates. They are related by the Taylor expansion
\begin{equation}
    f(\vx) = f( \vq + \Ps) = f(\vq) + \Psi_i(\vq) f_{,i}(\vq) + \frac{1}{2}\Psi_i(\vq) \Psi_j(\vq) f_{,ij}(\vq) + \cdots
\end{equation}
whose $q-$Fourier transform yields
\begin{align} \label{tildef}
\tilde{f}(\vk) &= \int d^3q\, e^{-i\vk \cdot \vq} f(\vx) = \int d^3q\, e^{-i\vk \cdot \vq} f(\vq + \Ps(\vq)) \\
               &= \int d^3q\, e^{-i\vk \cdot \vq} \Big( f(\vq) + \Psi_i(\vq) f_{,i}(\vq) + \frac{1}{2}\Psi_i(\vq) \Psi_j(\vq) f_{,ij}(\vq) + \cdots \Big) \nonumber \\
               &= f(\vk) + \ikk i k_1^i f(\vk_1) \Psi_i(\vk_2)    - \ikkk \frac{1}{2}k_1^i k_1^j f(\vk_1) \Psi_{i}(\vk_2)\Psi_{j}(\vk_3) + \cdots, \\
               &= f(\vk) +  \frac{i^n}{n!} \sum_{n=2}^\infty \underset{\vk_{1\cdots n}= \vk}{\int} k^{i_2}_1\cdots k^{i_n}_1 f(\vk_1) \Psi_{i_2}(\vk_2)\cdots  \Psi_{i_n}(\vk_n) 
\end{align}
and the inverse relation \cite{Aviles:2020cax}
\begin{align}\label{tftof}
 f(\vk)         &= \tilde{f}(\vk) -  \ikk  i k_1^i \tilde{f}(\vk_1) \Psi_i(\vk_2)    \nonumber\\
                &\quad  
                   + \ikkk \left( \frac{1}{2}k_1^i k_1^j -  k_{13}^ik_{1}^j\right) \tilde{f}(\vk_1) \Psi_{i}(\vk_2)\Psi_{j}(\vk_3) +\cdots,
\end{align}

We will use the above equations to write eq.~\eqref{eom} in terms of Lagrangian coordinates only. But before that, in the next subsection we will revisit the standard approach used in $\Lambda$CDM, where the linear theory is scale independent. 

\end{subsection}

\begin{subsection}{Standard approach in vanilla $\Lambda$CDM} \label{subsect:VanillaLCDM}
Before delving into more complex scenarios involving additional scales that arise when we incorporate the source term in eq.~\eqref{Poisson}, let us revisit the case of the vanilla $\Lambda$CDM model. To achieve this, we take the divergence of the equation of motion \eqref{eom} and utilize the Poisson equation (eq.~\eqref{Poisson} with $S=0$), yielding
\begin{equation}\label{eomVL}
 \Big[\vec \nabla_{\vx} \cdot \T \Ps(\vq)\Big](\vk) = - 4 \pi G \rho_m \tilde{\delta}(\vk).
\end{equation}
Our objective is to write this equation in terms of pure Lagrangian coordinates, so that we can subsequently expand it perturbatively in terms of the displacement $\Ps$. 
Using the Jacobian transformation matrix, 
\begin{equation}
    J_{ij}(\vq,t) \equiv \frac{\partial x_i(\vq,t)}{\partial q^j } = \delta_{ij} + \Psi_{i,j}(\vq,t), 
\end{equation}
where a comma means partial derivative with respect to Lagrangian coordinates, we obtain the transformation rule between spatial derivatives
\begin{equation}
    \nabla_{\vx\, i} = (J^{-1})_{ji} \frac{\partial \,}{\partial q^j}
\end{equation}
with $(J^{-1})_{ji}$ the inverse of the Jacobian matrix. 
The LHS of \eqref{eomVL} becomes 
\begin{equation}
 \vec \nabla_{\vx} \cdot \T \, \Ps(\vq) = (J^{-1})_{ji} \T  \Psi_{i,j} = \T \Psi_{i,i} - \Psi_{i,j} \T \Psi_{j,i} +  \Psi_{i,k} \Psi_{k,j} \T \Psi_{i,j} + \mathcal{O}(\Ps^4),  
\end{equation}
up to cubic terms in the Lagrangian displacement. 

Now, for the RHS of eq.~\eqref{eomVL} we use the matter conservation
\begin{equation} \label{matterconserv}
    \big(1+\delta(\vx,t) \big) d^3 x = d^3 q 
    = [J(\vq,t)]^{-1} d^3 x, 
\end{equation}
where $J = \det \big( J_{ij} \big)$ is the determinant of the Jacobian matrix, and the evolution is thought to be initiated sufficient early so we can neglect the Lagrangian fluctuation, $1+\delta(\vq) \approx 1$.  Then, one is able to relate the density field with the Lagrangian displacements
\begin{align} \label{deltainPsi}
    -\delta(\vx,t) = 1 -\frac{1}{J(\vq,t)} 
    &= \Psi_{i,i} - \frac{1}{2} \big((\Psi_{i,i})^2 + \Psi_{i,j}\Psi_{j,i}\big) \nonumber\\
    &\quad + \frac{1}{6}(\Psi_{i,i})^3 
    + \frac{1}{3}\Psi_{i,j}\Psi_{j,k}\Psi_{k,i} -\frac{1}{2}\Psi_{k,k}\Psi_{i,j}\Psi_{j,i} + \mathcal{O}(\Ps^4),
\end{align}
which is again written in pure Lagrangian coordinates.

Now, we can put all the ingredients together and write
\begin{align} \label{eqmVL}
 \big(\T - A_0 \big) [\Psi_{i,i}](\vk) &= [\Psi_{i,j} \T \Psi_{j,i}](\vk) 
 - \frac{A_0}{2}[\Psi_{i,j} \Psi_{j,i}](\vk) - \frac{A_0}{2} [(\Psi_{l,l})^2](\vk)  \nonumber\\
 &\quad  - [\Psi_{i,k}\Psi_{k,j} \T \Psi_{j,i}](\vk) +\frac{A_0}{6} [(\Psi_{l,l})^3](\vk) \nonumber\\
 &\quad 
  + \frac{A_0}{2} [\Psi_{l,l} \Psi_{i,j} \Psi_{j,i}](\vk) + \frac{A_0}{3} [\Psi_{i,k}\Psi_{k,j} \Psi_{j,i}](\vk), 
\end{align}
where we defined 
\begin{equation}
    A_0(t) \equiv 4 \pi G \rho_m  = \frac{3}{2} \Omega_m(a) H^2.
\end{equation}

We notice eq.~\eqref{eqmVL} is not the standard form to write the equation of motion, e.g. see \cite{Matsubara:2015ipa}. Indeed, it is notational simpler to omit the symbols $[\, \cdots \,](\vk)$. However,  it is better to keep the above form for later comparison when including the source $S(\vx)$ in the next subsection.

Because of the conservation equation \eqref{matterconserv}, we had no need to use eq.~\eqref{tildef} for the density field $\tilde{\delta}(\vk)$. This happens obviously for the vanilla $\Lambda$CDM model only, and hence the fact that one uses Fourier transforms in different frames  is rarely noticed in the literature. However, below we will be forced to use eq.~\eqref{tildef} when we include the source $S$.

\end{subsection}

\begin{subsection}{Source}\label{subsect:source}

Now, let us consider the source term in the Poisson equation. We assume that $S(\vk)$ has an expansion in the density fluctuations, that in Fourier space takes the form
\begin{equation}
S(\vk,t) = A_0(t) \alpha(k,t) \delta(\vk,t) + S^\text{NL}(\vk,t),
\end{equation}
where $A_0 \alpha(k)$ is the factor of proportionality of the linear term and $ S^\text{NL}(\vk,t)$ encapsulates the non-linearities. %

Adding the source to the density field as appearing in the Poisson equation, 
\begin{align*}
    A_0 \delta(\vx) + S(\vx)  
    \quad  
    \xrightarrow{\text{$\vx$-FT }}
    \quad 
      A_0  \delta(\vk) + S(\vk)   =&  A_0 \big(1+\alpha(k) \big) \delta(\vk)  + S^\text{NL}(\vk) \\
                                =&  A(k) \delta(\vk) + S^\text{NL}(\vk), 
\end{align*}
where we performed the $x$-Fourier transform and in the last line we defined the function
\begin{equation}
    A(k,t) = A_0(t) \big(1+\alpha(k,t) \big). 
\end{equation}

Now, based on eq.~\eqref{tildef}, the function $\alpha(k)$ in Eulerian space differs from its counterpart in Lagrangian space at first order in the relevant fields, in this case the Lagrangian displacement. Hence the combinations $\alpha(k) \delta(\vk)$ and $\alpha(k) \tilde{\delta}(\vk)$ differ only on non-linear terms in the density fluctuation. We refer to these non-linear terms as Frame-Lagging (FL).\footnote{The name \textit{frame-lagging} comes from ref.~\cite{Aviles:2017aor}, where the analogous terms in the specific case of MG arise when transform derivatives in the Klein-Gordon equation between the Eulerian and Lagrangian frames, showing the \textit{lag} of $q$-coordinates with respect to $x$-coordinates. In this case the \textit{lag} arises more generally since $f(\vx) = f(\vq) + \Psi_i f_i(\vq) + \cdots$. }  This property will allow us to obtain the function $\alpha(k)$ in Eulerian space where the physical phenomena are commonly described. For example, in massive neutrino cosmologies it is constructed with transfer functions coming out from a Boltzmann code \cite{Aviles:2020cax}. Thus, an important observation when transitioning to Lagrangian space is that we can retain the functional form of $\alpha(k)$ obtained in Eulerian space and add the corrections through FL terms. This process can be accounted for as follows.

For a Lagrangian space framework, we take the $q$-Fourier Transform (FT) 
\begin{equation}
 \qquad \, A_0\delta(\vx) + S(\vx) 
    \quad  
    \xrightarrow{\text{$\vq$-FT }}
    \quad 
   A_0\tilde{\delta}(\vk) + \tilde{S}(\vk)
\end{equation}

We now look for an expression that schematically is written as
\begin{displaymath}
\tilde{S}(\vk) = A_0 \alpha(k)\tilde{\delta}(\vk) + \text{Frame Lagging (FL)}  +  \tilde{S}^\text{NL}(\vk)
\end{displaymath}

Now, we omit for the moment the non-linear source $\tilde{S}^\text{NL}$, which typically carries the screening effect, and develop up to second order, 
\begin{align*}
\frac{\tilde{S}(\vk)}{A_0} \quad &= \quad \frac{S(\vk)}{A_0} + \int_{\vk=\vk_{12}} ik_1^i \frac{S(\vk_1)}{A_0} \,\Psi_i(\vk_2) + \cdots  
               \\
              &= \quad \alpha(k) \tilde\delta(\vk)  + \alpha(k) (\delta(\vk) - \tilde\delta(\vk)) \nonumber \\
              &\quad \quad + \int_{\vk=\vk_{12}} ik_1^i \Big[ \alpha(k_1) \tilde \delta(\vk_1) + \, \alpha(k_1) (\delta(\vk_1) - \tilde\delta(\vk_1)) \Big]\Psi_i(\vk_2) + \cdots \\ 
              %
              %
              &= \quad \alpha(k) \tilde \delta(\vk) - \int_{\vk=\vk_{12}} ik_i \Big[\alpha(k) - \alpha(k_1) \Big] \tilde{\delta}(\vk_1) \Psi_i(\vk_2)  \,\, + \,\, \cdots
\end{align*}
In the first equality we have expanded $\tilde{S}(\vk)$ into $S(\vk)$ using eq.~\eqref{tildef}. Then,  in the second equality, we have written $S/A_0 = \alpha \delta =\alpha \tilde{\delta} + \alpha (\delta - \tilde{\delta})$.   In the last equality 
we have used eq.~\eqref{tftof} and wrote $\delta(\vk) -  \tilde\delta(\vk) =  \int_{\vk=\vk_{12}}  (-ik_1^i) \tilde{\delta}(\vk_1) \Psi_i(\vk_2) \,\, + \cdots$, and finally grouped second order terms into the integral. Notice also that the term $(\delta - \tilde\delta)\Psi_i $ is third order and has been dropped out of this calculation.

The equation above still retains Eulerian coordinate dependencies through $\tilde{\delta}$, which we can eliminate in favor of the Lagrangian displacement using eq.~\eqref{deltainPsi}, leading to 
\begin{align}\label{sourceSFL}
 \frac{\tilde{S}(\vk)}{A_0}  &= \alpha(k) \tilde{\delta}(\vk) 
         + \ikk  \mathcal{K}^\text{FL}_{ki}(\vk_1,\vk_2) \Psi_k(\vk_1) \Psi_i(\vk_2)  \nonumber\\
&\quad   + \ikkk  \mathcal{K}^\text{FL}_{kij}(\vk_1,\vk_2,\vk_3) \Psi_k(\vk_1) \Psi_{i}(\vk_2)\Psi_{j}(\vk_3) + \cdots, 
\end{align}
with frame-lagging kernels
\begin{align}
 \mathcal{K}^\text{FL}_{ki}(\vk_1,\vk_2) &=  \big(\alpha(k_1)-\alpha(k)\big)  k_1^k k_1^i, \\
 \mathcal{K}^\text{FL}_{kij}(\vk_1,\vk_2,\vk_3) &= -i \big(\alpha(k)-\alpha(k_1) \big) k_1^k k_1^i k_1^j \nonumber\\
 &\quad    +  i \big( \alpha(k) - \alpha(k_{13}) \big) (k_1^i+k_3^i) \left( k_1^jk_1^k +\frac{1}{2} k_1^k k_3^j + \frac{1}{2}k_1^j k_3^k \right).            
\end{align}
where we used $\tilde\delta(\vk)  = -ik_i \Psi_i(\vk) + \cdots$ from  eq.~\eqref{deltainPsi}. Equation \eqref{sourceSFL} give the source in the Poisson equation \eqref{Poisson} transformed to the Lagrangian frame, with the peculiarity of having the function $\alpha(k)$ the same functional form as in Eulerian coordinates.    

\end{subsection}

\begin{subsection}{Lagrangian displacement evolution equation}\label{subsect:Lag_Disp}
 
 Finally, putting the above ingredients together, the geodesic equation becomes
\begin{align} \label{eqm}
 &\big(\T - A(k) \big) [\Psi_{i,i}](\vk) = [\Psi_{i,j} \T \Psi_{j,i}](\vk) 
 - \frac{A(k)}{2}[\Psi_{i,j} \Psi_{j,i}](\vk) - \frac{A(k)}{2} [(\Psi_{l,l})^2](\vk)  \nonumber\\*
 &\quad - [\Psi_{i,k}\Psi_{k,j} \T \Psi_{j,i}](\vk) +\frac{A(k)}{6} [(\Psi_{l,l})^3](\vk) 
  + \frac{A(k)}{2} [\Psi_{l,l} \Psi_{i,j} \Psi_{j,i}](\vk) + \frac{A(k)}{3} [\Psi_{i,k}\Psi_{k,j} \Psi_{j,i}](\vk) \nonumber\\*
  &\quad 
-  \ikk  \mathcal{K}^\text{FL$\Psi$}_{ki}(\vk_1,\vk_2) \Psi_k(\vk_1) \Psi_i(\vk_2) 
-   \ikkk  \mathcal{K}^\text{FL$\Psi$}_{kij}(\vk_1,\vk_2,\vk_3) \Psi_k(\vk_1) \Psi_{i}(\vk_2)\Psi_{j}(\vk_3),
\end{align}
which is the equation of motion written in terms of pure Lagrangian coordinates.

To solve this equation perturbatively, we express the Lagrangian displacements as a series expansion $\Ps =  \Ps^{(1)} +  \Ps^{(2)} + \Ps^{(3)} + \cdots $. Each term in the expansion can be further expanded as
\begin{equation}
\Psi^{(n)}_i =  \frac{i}{n!} \underset{\vk_{1\cdots n}= \vk}{\int} L^{(n)}_i(\vk_1,\cdots,\vk_n;t) \delta^{(1)}(\vk_1,t)\cdots\delta^{(1)}(\vk_n,t)      
\end{equation} 
Here, $\delta^{(1)}$ represents the first-order linear overdensity in Eulerian space. The integral involves a specific configuration of wavevectors $\vk_1,\cdots,\vk_n$, and $L^{(n)}_i(\mathbf{k}_1,\cdots,\mathbf{k}_n;t)$ is the kernel associated with the Lagrangian displacement at the $n$-th order. It is important to note that the kernel may have a time dependence, which is not considered in the standard approach that assumes GR and the relation $\Omega_m=f^2$.

\begin{subsubsection}{Linear order}
    After linearizing, eq.~\eqref{eqm} can be expressed as $\big(\T - A(k) \big) [\Psi_{i,i}](\vk) = 0$. This equation bears resemblance to the equation for the $\Lambda$CDM linear growth function $D_+$, but modulated by the $k$ dependence. Additionally, from eq.~\eqref{deltainPsi} one can confirm the relation between linear density and displacement fields is the same as in $\Lambda$CDM: 
\begin{equation}
    ik_i \Psi_i^{(1)}(\vk,t) = \delta^{(1)}(\vk,t) = \delta^{(1)}(\vk,t_0) D_+(k,t),
\end{equation}
where the linear growth function $D_+$ is the fastest growing solution to
\begin{equation}\label{Dplus}
   \big(\T - A(k,t) \big)  D_+(k,t) = 0, 
\end{equation}
which we normalize to unity at large scales and present time: $D_+(k=0,t_0)=1$. 

Therefore, the modification to the linear displacement arises solely from the extra $k$-dependence introduced by the linear growth function, which is also present in the linear density field. As a result, the first-order Lagrangian Perturbation Theory (LPT) kernel can be expressed as
\begin{equation}
    L^{(1)}_i(\vk) = i\frac{k_i}{k^2}, 
\end{equation}
as in the $\Lambda$CDM model. 

\end{subsubsection}

\begin{subsubsection}{2LPT}

Keeping up to second order displacements in eq.~\eqref{eqm}, we obtain the second order kernel \cite{Aviles:2017aor,Aviles:2020cax} 
\begin{equation}
    L^{(2)}_i(\vk_1,\vk_2,t) = i \frac{k_i}{k^2} \left[ \mA(\vk_1,\vk_2,t) - \mB(\vk_1,\vk_2,t) \left( \frac{\vk_1 \cdot \vk_2}{k_1 k_2}\right)^2 \right] 
\end{equation}
with $\vk = \vk_1 + \vk_2$. The functions $\mA$ and $\mB$ are obtained by first solving the equations 
\begin{align}
\big(\T - A(k)\big) D^{(2)}_{\mA} &= \Bigg[A(k)   + (A(k)-A(k_1))\frac{\vk_1\cdot\vk_2}{k_2^2}  + (A(k)-A(k_2))\frac{\vk_1\cdot\vk_2}{k_1^2} 
             \nonumber\\
             &\quad  - S_2(\vk_1,\vk_2) \Bigg]  D_{+}(k_1)D_{+}(k_2), \label{DAeveq} \\
\big(\T - A(k)\big) D^{(2)}_{\mB} &= \Big[A(k_1) + A(k_2) - A(k) \Big] D_{+}(k_1)D_{+}(k_2), \label{DBeveq}
\end{align}
subject to appropriate initial conditions to isolate the pure second order contributions, and  
\begin{align} \label{AandBdef}
 \mA(\vk_1,\vk_2,t) &= \frac{7 D^{(2)}_{\mA}(\vk_1,\vk_2,t)}{3 D_{+}(k_1,t)D_{+}(k_2,t)}, \nonumber\\
 \mB(\vk_1,\vk_2,t) &= \frac{7 D^{(2)}_{\mB}(\vk_1,\vk_2,t)}{3 D_{+}(k_1,t)D_{+}(k_2,t)},
\end{align}

If $A(k,t) = A(t)=A_0(1+\alpha(t))$ is scale independent, both functions are equal $\mA=\mB$. 
In EdS background evolution we recover $\mA(t) = \mB(t)= 1$; whereas for the $\Lambda$CDM model, $\mA(t) = \mB(t)$ are only weakly dependent on time and close to one. For standard cosmologies one finds that nowadays $\mA^{\Lambda\text{CDM}}(t_0) \simeq 1.01$.

The term $S_2(\vk_1,\vk_2)$ in Eq.~(\ref{DAeveq}) arises from the non-linear source term $S_\text{NL}$ and is responsible for the screening mechanism to second order in perturbation theory. For the HS-$f(R)$ model this is given by
\begin{align}\label{sourceS}
S^\text{HS}_2(\vk_1,\vk_2) =    \frac{36 \Omega_m^2 H^4 \beta^6 a^4 M_2(\vk_1,\vk_2)k^2}{ (k^2 + m^2a^2)(k_1^2 + m^2a^2)(k_2^2 + m^2a^2)},
\end{align} 
with $M_2$ given by eq.~\eqref{M2}.
As will be discussed in sec.~\ref{sect:ScreeningVsEFT}, these terms are degenerate with EFT counterterms and will not be considered in this work. An alternative scenario that considers the screenings, and promises to be fast, was recently given in \cite{Euclid:2023bgs}. But, we emphasize that one of the objectives of this paper is to show that screenings can be avoided in some cases because of their degeneracies with EFT parameters.

Now, in one-loop integrals only the longitudinal piece of the displacement field shows up \cite{Matsubara:2007wj,Carlson_2012}, so we can define the kernels $\Gamma_n$ through \cite{Valogiannis:2019nfz}
\begin{equation}\label{LDDkernels}
 -i\vk \cdot \Ps^{(n)}(\vk,t) = \frac{1}{n!} \underset{\vk_{1\cdots n}= \vk}{\int} C_{n} \Gamma_n(\vk_1,\cdots,\vk_n,t)\delta^{(1)}(\vk_1,t)\cdots \delta^{(1)}(\vk_n,t),
\end{equation}
where $C_n$ are constant coefficients that we fixed to the values $C_1=C_3=1$, and $C_2 = 3/7$. $\Gamma_n$ and $L_i^{(n)}$ kernels are related by 
\begin{equation}
 C_n \Gamma_n(\vk_1,\cdots,\vk_n,t) = (k_1^i + \cdots +k_n^i) L_i^{(n)}(\vk_1,\cdots,\vk_n;t).
\end{equation}
A rapid computation yields the kernels $\Gamma_1(\vk)=1$ to first order, and
\begin{align} \label{Gamma2}
\Gamma_2(\vk_1,\vk_2,t) &= \mA(\vk_1,\vk_2,t) - \mB(\vk_1,\vk_2) \frac{(\vk_1 \cdot \vk_2)^2}{k_1^2 k_2^2}
\end{align}
to second order. 
The third order kernels can be found in Appendix \ref{app:3rdOrder}.

\end{subsubsection}

\begin{subsubsection}{Velocity fields}

For longitudinal flows, only the divergence of the velocity field is significant, hence it is useful to define 
\begin{equation}\label{deftheta}
    \theta(\vk) \equiv -\frac{i \vk \cdot \vec v}{aH f_0} = - \frac{i \vk \cdot \dot{\Ps}}{H f_0},
\end{equation}
where the peculiar velocity of particles relative to the Hubble flow is given in terms of the Lagrangian displacement time derivative by $\vv = a \dot{\Ps}$. We also defined $f_0$ as the large scale limit of the growth rate, that is,
\begin{equation}
    f(k,t) \equiv \frac{d \log D_+(k,t)}{d\log a(t)}, \qquad f_0(t) = f(k=0,t). 
\end{equation}

Using eqs.~\eqref{LDDkernels} and \eqref{deftheta} we have 
\begin{align}
 \theta(\vk) &=   \sum_{m=1}^{\infty} \frac{1}{(m-1)!} \underset{\vk_{1\cdots m} = \vk}{\int}  \Gamma^f_{m}(\vk_1,\dots,\vk_m,t)
 \delta^{(1)}(\vk_1,t) \cdots \delta^{(1)}(\vk_{n},t), 
\end{align}
where
\begin{equation}\label{LDDfkernels}
 \Gamma^f_n(\vk_1,\cdots,\vk_n,t) \equiv  \Gamma_n(\vk_1,\cdots,\vk_n,t) \frac{f(k_1)+\cdots+f(k_n)}{n f_0} + \frac{1}{n f_0 H} \dot{\Gamma}_n(\vk_1,\cdots,\vk_n,t).
\end{equation}

In the EdS case, we obtain $\Gamma^f_n=\Gamma_n$, recovering the well-known relation $\dot{\Ps}^{(n)} = n H f \Ps^{(n)}$ \cite{Matsubara:2007wj}. 
At first order, we have  $\Gamma_1^f(\vk)=f(k)/f_0$, which leads to the linear relationship between the velocity and density fields
\begin{equation}\label{theta1delta1}
    \theta^{(1)}(\vk,t) =  \frac{f(k,t)}{f_0(t)}\delta^{(1)}(\vk,t).  
\end{equation}
This relation will play a central role in the subsequent analysis.

\end{subsubsection}

\end{subsection}

\begin{subsection}{Map to SPT Kernels} \label{subsect:LPTtoSPT}

An alternative and more widely used path to cosmological Large Scale Structure Perturbation Theory is given by the Eulerian SPT formalism \cite{Bernardeau:2001qr}. Here, one expands the fields $\theta=\theta^{(1)}+\theta^{(2)} + \cdots$ and $\delta=\delta^{(1)} + \delta^{(2)} + \cdots$ with
\begin{align}
 \delta^{(n)}(\vk,t) &= \underset{\vk_{1\cdots n}= \vk}{\int} F_n(\vk_1,\cdots,\vk_n,t)\delta^{(1)}(\vk_1,t)\cdots \delta^{(1)}(\vk_n,t), \label{dcbExp}\\
 \theta^{(n)}(\vk,t) &= \underset{\vk_{1\cdots n}= \vk}{\int} G_n(\vk_1,\cdots,\vk_n,t)\delta^{(1)}(\vk_1,t)\cdots \delta^{(1)}(\vk_n,t), \label{tcbExp}
\end{align}

We can use eqs.~\eqref{deltainPsi}, \eqref{LDDkernels} and \eqref{LDDfkernels} to relate Lagrangian and Eulerian kernels. Following \cite{Matsubara:2011ck,Aviles:2018saf,Aviles:2020wme} we obtain to second order
\begin{align}
 F_2(\vk_1,\vk_2) &= \frac{3}{14} \Gamma_2(\vk_1,\vk_2) + \frac{1}{2}\frac{(\vk_{12}\cdot\vk_1)(\vk_{12}\cdot\vk_2)}{k_1^2 k_2^2},  \label{LPTtoF2} \\
G_2(\vk_1,\vk_2)  &= \frac{3}{7} \Gamma^f_2(\vk_1,\vk_2) + \frac{(\vk_1\cdot\vk_2)^2}{k_1^2 k_2^2} \frac{f(k_1)+f(k_2)}{2f_0} \nonumber\\
&\quad + \frac{1}{2}\frac{\vk_1\cdot\vk_2}{k_1 k_2}  \left( \frac{k_2}{k_1}\frac{f(k_2)}{f_0}+\frac{k_1}{k_2}\frac{f(k_1)}{f_0} \right), \label{LPTtoG2} 
\end{align}
For third order kernels the relation is more cumbersome. But it simplifies for the particular configuration of wavevectors used in one-loop integrals
\begin{align}
 F_3(\vk,-\vp,\vp) &= \frac{1}{6} \Gamma_3(\vk,-\vp,\vp) -\frac{1}{6} \frac{(\vk\cdot\vp)^2}{p^4}  
 -  \frac{1}{7}\frac{(\vk \cdot (\vk-\vp)) (\vk\cdot\vp)}{|\vk-\vp|^2 p^2}\Gamma_2(\vk,-\vp). \\
 G_3(\vk,-\vp,\vp) &=  
   \frac{1}{2} \Gamma^{f}_3(\vk,-\vp,\vp)   +  \frac{2}{7} \frac{\vk \cdot \vp}{p^2} \Gamma^{f}_2(\vk,-\vp)   + \frac{1}{7}  \frac{f(p)}{f_0}  \Gamma_2(\vk,-\vp) \frac{\vk \cdot (\vk-\vp)}{|\vk-\vp|^2} \nonumber\\
&\quad    -  \frac{1}{7}   \left[ 2 \Gamma_2^f(\vk,-\vp) + \Gamma_2(\vk,-\vp) \frac{f(p)}{f_0} \right]\left[1 -\frac{(\vp \cdot (\vk-\vp))^2}{p^2|\vk-\vp|^2}  \right]\nonumber\\
&\quad - \frac{1}{6} \frac{(\vk \cdot \vp)^2}{p^4} \frac{f(k)}{f_0}   .
\end{align}

\end{subsection}

\begin{subsection}{Power spectrum}\label{subsect:ps}
The apparent position $\vs$ of an object is distorted from its true position $\vx$ because of the Doppler effect induced by its peculiar velocity, such that we observe it at a redshift space coordinate
\begin{equation}\label{RSDmap}
 \vs(\vx) = \vx + \vu(\vx), \quad \text{with} \quad \vu(\vx) =  \int \Dk{k} e^{i\vk\cdot\vx}\, i f_0 \vhn \frac{\vk \cdot \vhn}{k^2} \theta(\vk)
\end{equation}
Since the map from real to redshift coordinates conserves the number of tracers, $\big[1+\delta_s(\vs)\big]d^3s = \big[1+\delta(\vx)\big]d^3x$, we have
\begin{equation}
 (2\pi)^3\delta_\text{D}(\vk) + \delta_s(\vk) = \int d^3x \big(1+\delta(\vx)\big) e^{-i \vk \cdot(\vx+ \vu(\vx))},  
\end{equation}
and the redshift-space PS becomes  \cite{Scoccimarro:2004tg,Vlah:2018ygt}
\begin{equation}\label{RSDPS}
  (2\pi)^3\delta_\text{D}(\vk) + P_s(\vk) = \int d^3x e^{-i\vk\cdot \vx} \Big[ 1+\mathcal{M}(\vec J= \vk,\vx) \Big], 
\end{equation}
with the velocity moments generating function  
\begin{equation}\label{VDgenF}
1+\mathcal{M}(\vec J,\vx) =  \left\langle \big(1+\delta(\vx_1)\big)\big(1+\delta(\vx_2)\big)  e^{-i \vec J \cdot \Delta \vu}  \right\rangle,
\end{equation}
where $\Delta \vu = \vu(\vx_2)-\vu(\vx_1)$ and $\vx = \vx_2 - \vx_1$. Function $\mathcal{M}$ (or its Fourier transform) plays a central role in RSD. Different expansion procedures of eq.~\eqref{VDgenF} yield different approaches to RSD modeling \cite{Vlah:2018ygt}. 
Our approach follows the moment expansion (ME) approach of \cite{Scoccimarro:2004tg}, that uses a Taylor expansion of the generating function.
That is, the $\tm$-th density weighted velocity field moment of the generating function is an $\tm$-rank tensor defined as  \cite{Scoccimarro:2004tg,Vlah:2018ygt}
\begin{align}
 \Xi^{\tm}_{ i_1 \cdots i_\tm}(\vx) &\equiv i^\tm \frac{\partial^\tm}{\partial J_{i_1}\cdots \partial J_{i_\tm}} \big[ 1+\mathcal{M}(\vec J,\vx) \big] \Big|_{\vec J = 0} \nonumber\\
 &= \langle \big(1+\delta(\vx_1)\big)\big(1+\delta(\vx_2)\big)\Delta u_{i_1}\cdots\Delta u_{i_\tm} \rangle,
\end{align}
with $\delta_1 =\delta(\vx_1)$ and $\delta_2 =\delta(\vx_2)$. Hence, from eq.~\eqref{RSDPS}, the power spectrum in the moment expansion approach becomes
\begin{align} \label{PSmomexp}
   (2\pi)^3 \dD(\vk) + P_s(\vk)
      &= \sum_{\tm=0}^\infty \frac{(-i)^\tm}{\tm!} k_{i_1}\dots k_{i_\tm}  \tilde{\Xi}_{i_{1}\cdots i_{\tm}}^{\tm}(\vk),
\end{align}
where the $\tilde{\Xi}^{\tm}_{i_{1}\cdots i_\tm}(\vk)$  are the Fourier moments of the generating function 
---the Fourier transforms of their configuration space counterparts, $\Xi^{\tm}_{i_{1}\cdots i_{n}}(\vx)$. Finally, one can write  \cite{Aviles:2020wme,Aviles:2021que}
\begin{equation}
  \frac{(-i)^\tm}{\tm!} k_{i_1}\cdots k_{i_\tm} \tilde{\Xi}^{\tm}_{i_1\cdots i_\tm}(\vk) = \sum_{n=0}^\tm \mu^{2n} f_0^\tm I^\tm_{n}(k)   
\end{equation}
for some functions $I^\tm_{n}(k)$. Hence, the momentum expansion redshift space power spectrum can be written as
\begin{equation}\label{PsInPm}
P_s^\text{ME}(k,\mu)=\sum_{\tm=0}^\infty  \sum_{n=0}^\tm \mu^{2n} f_0^\tm I^\tm_{n}(k),       
\end{equation}
up to a Dirac delta function localized at $\vk=0$. 

When accounting for EFT corrections and shot-noise, the expression for the one-loop power spectrum is given by
\begin{equation}\label{pofk2}
   P_s(k, \mu) = P_s^\text{ME}(k, \mu) + P_s^\text{EFT}(k, \mu) + P_s^\text{shot}(k, \mu),
\end{equation}
which is composed by the following elements:

\begin{enumerate}
    \item The momentum expansion perturbation theory power spectrum
\begin{equation}\label{PsME}
   P_s^\text{ME}(k, \mu) = P_{\delta\delta}(k) + 2 f_{0} \mu^2 P_{\delta\theta}(k) + f_{0}^2 \mu^4 P_{\theta\theta}(k) + A^\text{TNS}(k,\mu) + D(k,\mu),  
\end{equation}
where the one-loop real space power spectra $P_{\delta \delta}$, $P_{\delta\theta}$, and $P_{\theta\theta}$ are presented below in \S \ref{subsect:biasing}. The function $A^\text{TNS}$ is defined in \cite{Taruya:2010mx} as
\begin{align}\label{ATNS}
A^\text{TNS}(k,\mu) &= 2 k \mu f_0 \int \Dk{p}  \frac{\vp\cdot \vhn}{p^2} B_\sigma(\vp,-\vk,\vk-\vp) \, , 
\end{align}
with the bispectrum $B_\sigma$ given through
\begin{align}\label{defBsigma}
&(2 \pi)^3 \dD(\vk_1 + \vk_2 + \vk_3) B_\sigma(\vk_1,\vk_2,\vk_3) = \nonumber\\
&\qquad \qquad \Big\langle \theta(\vk_1)\left[\delta(\vk_2) + f_0 \frac{(\vk_2\cdot\vhn)^2}{k_2^2}\theta(\vk_2) \right] \left[\delta(\vk_3) + f_0 \frac{(\vk_3\cdot\vhn)^2}{k_3^2}\theta(\vk_3) \right]\Big\rangle.
\end{align}

Meanwhile, the function $D(k,\mu)$ is given by
\begin{align} \label{DTNS}
D(k,\mu) &= (k\mu f_0)^2  \int \Dk{p} \Big\{ F(\vp)F(\vk-\vp)  \nonumber\\
      &\quad+   \frac{(\vp\cdot\vhn)^2}{p^4}P_{\theta\theta}^L(p) \big[ P^K_s(|\vk-\vp|, \mu_{\vk-\vp}) -P^K_s(k, \mu) \big] \Big\}, 
\end{align}
with $\mu_{\vk-\vp}$ the cosine of the angle between the wave-vector $\vk-\vp$ and the line-of-sight direction $\vhn$, and function $F$ is given by  \cite{Taruya:2010mx}
\begin{align}
F(\vp) &= \frac{\vp\cdot\vhn}{p^2}\Big[ P_{\delta\theta}(p) + f_0 \frac{(\vp\cdot\vhn)^2}{p^2} P_{\theta\theta}(p) \Big].
\end{align}
While
\begin{equation}
    P^K_s(k, \mu) = \big( 1 + \mu^2 f(k) \big)^2 P_L(k)
\end{equation}
is the linear Kaiser power spectrum \cite{Kaiser:1984sw}, but with the additional $k$-dependence in the growth rate.

We notice that to linear order, we can use the relation between velocity and densities given in eq.~\eqref{theta1delta1} to write
\begin{equation}
    P_{\theta\delta}^{L} =   \frac{f(k)}{f_0} P_{\delta\delta}^{L} \quad \text{and} \quad
    P_{\theta\theta}^{L} =   \left(\frac{f(k)}{f_0}\right)^2 P_{\delta\delta}^{L}, 
\end{equation}
and further, since functions  $A^\text{TNS}$ and  $D$ are pure non-linear, 
one recovers the Kaiser power spectrum. In the following, we refer to the density-density linear power spectrum $P_{\delta\delta}^{L}$ simply as $P_{L}$.

\item The EFT counterterms
\begin{equation}
    P_s^\text{EFT} = \big(\alpha_0 + \alpha_2 \mu^2 + \alpha_4 \mu^4 + \alpha_6 \mu^6\big) k^2 P_L(k) + \tilde{c} \big(\mu k f_0 \big)^4   P^K_s (k,\mu)   
\end{equation}

\item The shot noise
\begin{equation}
    P_s^\text{shot}(k, \mu) = \frac{1}{\bar{n}} \big( \alpha_0^\text{shot} + (k \mu)^2 \alpha^\text{shot}_2 \big)
\end{equation}
with $\bar{n}$ the average number density of galaxies, such that for a Poissonian distribution $ \alpha_0^\text{shot}=1$ and  $ \alpha_2^\text{shot}=0$.

\end{enumerate}

Despite the success of SPT-EFT in modeling the broadband power spectrum, the theory yet gives poor results in modeling the BAO  since long-wavelength displacement fields, though being essentially linear, stream largely contributing to damp features in the power spectrum in a manner that is non-perturbative under an SPT scheme. Then, in order to model the spread and degradation of the BAO oscillations due to large scale bulk flows, we employ IR-resummations  \cite{Senatore:2014via} as implemented by Ivanov et al.  \cite{Ivanov:2018gjr,Ivanov:2019pdj}. Here, we split the linear power spectrum in a piece with the wiggles removed, $P_{nw}$, and a wiggles piece, $P_{w}$, such that the spectrum can be written as $P_L = P_{nw} + P_w$. Defining  $ P_s(k,\mu)$ as the non-resummed full-spectrum computed through eq.\eqref{pofk2} using the complete linear power spectrum $P_L$, and analogously $P_{s,nw}(k,\mu)$,  but using only the non-wiggle piece $P_{nw}$, the IR-resummed power spectrum is \cite{Ivanov:2018gjr}
\begin{align}\label{PsIR}
P_s^\text{IR}(k,\mu) &= 
 e^{-k^2 \Sigma^2_\text{tot}(k,\mu)} P_s(k,\mu) +  \big(1-e^{-k^2 \Sigma^2_\text{tot}(k,\mu)} \big) P_{s,nw}(k,\mu) \nonumber\\
 &\quad +  e^{-k^2 \Sigma^2_\text{tot}(k,\mu)} P_w(k) k^2 \Sigma^2_\text{tot}(k,\mu). 
\end{align}
with
\begin{equation}\label{Sigma2T}
\Sigma^2_\text{tot}(k,\mu) = \big[1+f \mu^2 \big( 2 + f \big) \big]\Sigma^2 + f^2 \mu^2 (\mu^2-1) \delta\Sigma^2,    
\end{equation}
and
\begin{align}
\Sigma^2 &= \frac{1}{6 \pi^2}\int_0^{k_s} dp \,P_{nw}(p) \left[ 1 - j_0\left(p \,\ell_\text{BAO}\right) + 2 j_2 \left(p \,\ell_\text{BAO}\right)\right],\\
\delta\Sigma^2 &= \frac{1}{2 \pi^2}\int_0^{k_s} dp \,P_{nw}(p)  j_2 \left(p \,\ell_\text{BAO}\right).
\end{align}
The BAO peak scale is roughly given by $\ell_\text{BAO}\simeq 105 \hmpc$, with $j_n$ the spherical Bessel function of degree $n$. The choice of the scale $k_s$ that splits between long and short modes is arbitrary, but the results are very robust for $k_s\gtrsim 0.1 \hmpci$. Our code 
\texttt{fkpt} uses the value $k_s = 0.4 \hmpci$.

Finally, to fit the data, we use the multipoles from the equation
\begin{equation}\label{Pells}
P_\ell(k) = \frac{2 \ell + 1}{2} \int_{-1}^{1} d\mu \, P_s^\text{IR}(k,\mu) \mathcal{L}_{\ell}(\mu),    
\end{equation}
where $\mathcal{L}_{\ell}$ is the Legendre polynomial of degree $\ell$.

\end{subsection}

\begin{subsection}{Biasing}\label{subsect:biasing}

It is well known that there is no complete biasing theory for general theories with linear scale-dependent growth (e.g. \cite{Desjacques:2016bnm}), and one must add higher order derivative bias operators of the form $\nabla^2 \delta, \dots$. However, these terms become degenerate with EFT counterterms. Hence, to describe the galaxy-matter connection, it suffices to use the EFT theory of bias of \cite{McDonald:2006mx,McDonald:2009dh}, with some tweaks studied in \cite{Aviles:2020wme}. We have the biased spectra   
    \begin{align}
 P_{\delta\delta}(k) &= b_1^2 P^\text{1-loop}_{m,\delta\delta}(k)  + 2 b_1 b_2 P_{b_1b_2}(k) + 2 b_1 b_{s^2} P_{b_1b_{s^2}}(k) + b_2^2 P_{b_2^2}(k)  \nonumber\\
                     &\quad + 2 b_2 b_{s^2} P_{b_2 b_{s^2}}(k) + b_{s^2}^2 P_{b_{s^2}^2}(k) + 2 b_1 b_{3nl} \sigma^2_3(k) P^L_{m,\delta\delta}(k),  \label{PddTNL}\\
 P_{\delta\theta}(k) &=  b_1 P^\text{1-loop}_{m,\delta\theta}(k) + b_2 P_{b_2,\theta}(k)  + b_{s^2} P_{b_{s^2},\theta}(k) + b_{3nl} \sigma^2_3(k)  P^L_{m,\delta\theta}(k), \label{PdtTNL}\\
 P_{\theta\theta}(k) &=   P^\text{1-loop}_{m,\theta\theta}(k),  \label{PttTNL}                
\end{align}
where the function $\sigma_3(k)$ is given by eq. (3.26) of \cite{Aviles:2020wme} and the biased spectra of the form $P_{XY}$ are given by eqs.(3.43)-(3.49) of the same reference. For the biased functions $A^\text{TNS}$ and $D(k,\mu;f)$ we use
\begin{align}
    A^\text{TNS}(k,\mu;f) = b_1^3 A^\text{TNS}_m(k,\mu;f_{0}/b_1),  \\
    D(k,\mu;f) = b_1^4 D_m(k,\mu;f_{0}/b_1). 
\end{align}

\end{subsection}

\end{section}

\begin{section}{fk-perturbation theory} \label{sec:fkPerturbationTheory}

In the following, we describe the fk-Perturbation Theory (fkPT), which approximates the full theory described in the previous section. We further present its implementation in a fast C code, \texttt{fkpt}, that allows us to sample a large space of parameters with standard MCMC recipes.

\begin{subsection}{fk-kernels} \label{subsec:fkkernels}
The power spectrum can be written as a sum of $k$-functions multiplied by powers of the linear growth and the cosine angle $\mu$
\begin{equation}
    P(k,\mu) = \sum_{m}\sum_{n} \mu^{2n} f^m_0 I_{mn}(k)
\end{equation}
with
\begin{equation}
    I_{mn}(k) = \int d \vp \, \mathcal{I}(\vk,\vp)
\end{equation}
and the functions $\mathcal{I}(\vk,\vp)$ are invariant under spatial rotations. That is they only depend on the magnitudes $p = |\vp|$ and $k = |\vk|$ and on the angle between $\vp$ and $\vq$, and therefore the functions $I$ are indeed only functions of the wave-vector magnitude $k$. There are two kinds of these functions 
\begin{align}
      I_{mn}(k) = \int \Dk{p}  \mathcal{K}(\vp,\vk-\vp) P_L(p) P_L(|\vk-\vp|), \qquad & \text{(type $P_{22}$)}, \\
      I_{mn}(k) = P_L(k) \int \Dk{p}  \mathcal{K}(\vk,\vp)  P_L(\vp), \qquad & \text{(type $P_{13}$)}.
\end{align}
When using EdS evolution, or more precisely when $\Omega_m^2 = f$, one has explicit analytical expressions for the kernels $\mathcal{K}$. However, for theories that introduce new scales one has to solve differential equations for obtaining its precise form. These equations depend on the cosmological parameters, but also on the wave-vectors. For example, one of the $I$ functions is 
\begin{align}
    P_{22}^{\theta\theta}(k) = \int \Dk{p} \big[G_2(\vp,\vk-\vp)\big]^2 P_L(p) P_L(|\vk-\vp|) 
\end{align}
with (see eq.~\eqref{LPTtoG2})  
\begin{align}
G_2(\vk_1,\vk_2) &=    \frac{3\,\big[f(k_1)+f(k_2)\big]\,\mathcal{A} + 3\dot{\mathcal{A}}/H }{14 f_0}  + \frac{\hat{\vk}_1\cdot\hat{\vk}_2}{2 } \left( \frac{f(k_2)}{f_0}\frac{k_2}{k_1} + \frac{f(k_1)}{f_0} \frac{k_1}{k_2} \right) \nonumber\\
&\quad + \Big(\hat{\vk}_1\cdot\hat{\vk}_2\Big)^2
\left(\frac{f(k_1)+f(k_2)}{2 f_0} - \frac{3 \,\big[f(k_1)+f(k_2)\big] \,\mathcal{B} + 3\dot{\mathcal{B}}/H }{14 f_0}\right). 
\end{align}
This kernel will serve us to illustrate our approach:
We notice, there are two new types of contributions in the $G_{22}$ kernel that are not present in the EdS case: 
\begin{enumerate} 
    \item The first type comes from the linear growth rates inherited from linear theory. In particular, the dipole term is determined by the advection of density fields. As discussed in \cite{Aviles:2021que},  the second-order velocity field has a term given by
\begin{equation}
\theta^{(1)}(\vx+ \Psi) -\theta^{(1)}(\vx) 
= \partial_i\big[\nabla^{-2}\delta^{(1)}(\vx)\big]\partial_i \theta^{(1)}(\vx) \in \theta^{(2)}(\vx)   
\end{equation}
and in virtue of eq.~\eqref{theta1delta1},
\begin{equation} \label{gradientG2}
\frac{\hat{\vk}_1 \cdot \hat{\vk}_2}{2} \left(\frac{f(k_2)}{f_0}\frac{k_2}{k_1} + \frac{f(k_1)}{f_0}\frac{k_1}{k_2}\right) \in G_2(\vk_1, \vk_2).    
\end{equation}  

    \item The second type of contribution arises from the presence of the functions $\mathcal{A}(\vk_1,\vk_2,t)$, $\mathcal{B}(\vk_1,\vk_2,t)$, and their derivatives. These functions appear when the equation $f^2=\Omega_m$ fails to hold true, and, in the case of MG, they carry the screening effects.
\end{enumerate}

It turns out that the dominant corrections to the EdS kernels are of the first type. In the case of $f(R)$ theories, the corrections to $G_{22}$ induced by the functions $\mathcal{A}$ and $\mathcal{B}$ are about 1-2\% \cite{Aviles:2017aor}. 
Indeed, the lack of precise numerical values for these terms is not much more harmful than the use of EdS kernels when fitting the $\Lambda$CDM, which is a method that has proven to be quite accurate when fitting simulations and real data. On the other hand, the corrections provided by the growth rates can exhibit significantly larger magnitudes. To illustrate this, in figure \ref{fig:fk} we plot the function $f(k)/f_0$ for the scenarios F4, F5, and F6 at redshifts $z=0.5$ and $z=1.5$. This shows that the corrections to $G_{22}$ can surpass the $15\%$.

 \begin{figure}
 	\begin{center}
 	\includegraphics[width=3.5 in]{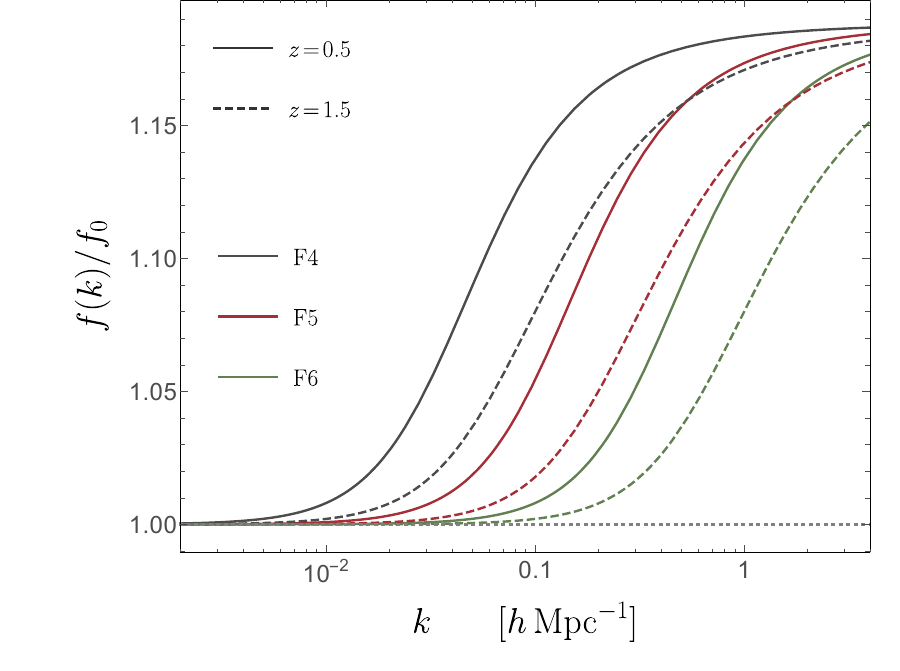}
 	\caption{Linear growth function $f$ as a function of $k$ for different redshifts ($z=0$ and $z=1.5$) and different gravitational strengths (F4, F5 and F6). All cases are normalized to one at large scales by showing $f(k,z)/f_0(z)$. } 
 	\label{fig:fk}
 	\end{center}
 \end{figure}

The above arguments point out that the use of EdS kernels to model the LSS in scale-dependent MG may not be correct. How bad is this method, clearly depends on the particular MG model. On the other hand, the use of the precise exact kernels is not computationally viable for an MCMC analysis. Hence we adopt the use of \texttt{fk}-kernels, which were introduced for massive neutrino cosmologies in \cite{Aviles:2021que} and further developed in \cite{Noriega:2022nhf}. The \texttt{fk}-kernels, which consider the exact $f(k,t)$ functions, can be defined as
\begin{align*}
F_2^{\texttt{fk}}(\vk_1,\vk_2) &=  F_2(\vk_1,\vk_2)\Big|_{\mathcal{A}=\mathcal{B}=\mathcal{A}^\text{LS}},   \\
G_2^{\texttt{fk}}(\vk_1,\vk_2) &=  G_2(\vk_1,\vk_2)\Big|_{\mathcal{A}=\mathcal{B}=\mathcal{B}^\text{LS}}, 
\end{align*} 
and similar for $n>2$.  That is, in this approach one fixes functions $\mathcal{A}$ and $\mathcal{B}$ to their large scale counterparts $\mathcal{A}^\text{LS}(t)$ and $\mathcal{B}^\text{LS}(t)$, obtained by evaluating all momenta at zero value. By eliminating the $k$-dependence from the function $A(k,\mu)$ in eqs.~\eqref{DAeveq} and \eqref{DBeveq} we observe that the large scale values of $\mA$ and $\mB$ are equal when the screening vanishes. This behavior is expected in theories that converge to GR for large scales, such as scale-dependent theories like $f(R)$, but not in theories with a massless scalar field mediator, such as DGP.

One can set the value of $\mathcal{A}^\text{LS}$ and $\mathcal{B}^\text{LS}$ to unity as in EdS, which could be a good idea for the cases in which the additional gravitational scalar degree of freedom is massive and hence the associated fifth force has a finite range, as in $f(R)$. However, preserving their exact large-scale values does not significantly impact computational time, as the differential equations only need to be solved once in the limit where all momenta tend to zero. This approach may be particularly beneficial, if not necessary, in theories that do not converge to the standard $\Lambda$CDM model at large scales, such as DGP and cubic Galileons. For the specific case of the scale-independent normal branch of DGP, ref.~\cite{Piga:2022mge} uses the correct large scale values of the kernels and find good agreement when confronting to simulations. 

Throughout this work we will use the exact large scales values of the functions $\mathcal{A}^\text{LS}$, $\mathcal{B}^\text{LS}$, ..., instead of the ones in EdS where these functions are unity. The latter scheme assumes the approximation $\Omega_m=f^2$ to be valid at large scales, and we call it EdS-\texttt{fk}. Figure \ref{fig:ScreeningsVsEFT} shows the power spectrum ratios for EdS-\texttt{fk} and \texttt{fk} kernels for multipoles $\ell=0$ and 2 (black and red dotted lines) for the model F5 at redshift $z=0.38$. It shows that the difference is smaller than around 1\% in the range of interest for full-shape analyses.

 \begin{figure}
 	\begin{center}
 	\includegraphics[width=3.5 in]{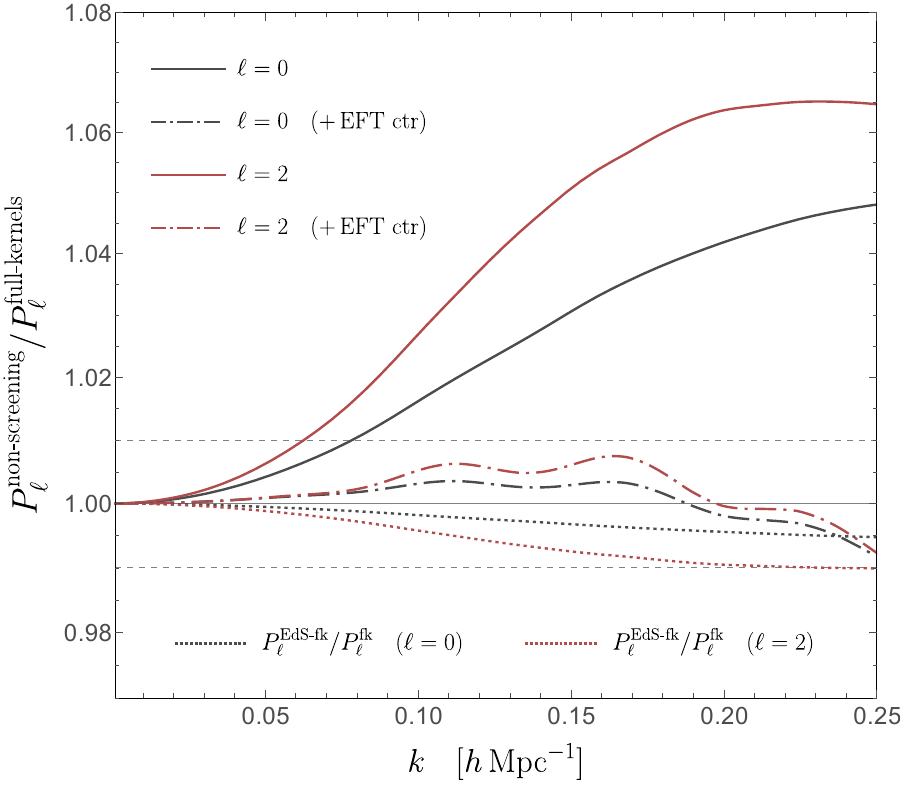}
 	\caption{We show the screenings-EFT counterterms degeneracy by taking the ratio of the power spectrum multipoles $\ell= 0$ and 2 without screenings and considering full kernels (solid lines). In dot-dashed lines we show the results of adding counterterms of the form $\alpha k^2 P_L(k)$ to $P_0$ and to $P_2$ produce similar effects than the screenings. We also show the ratio of the power spectrum multipoles when using EdS-\texttt{fk} and \texttt{fk}-kernels (dotted lines), as explained in \S \ref{subsec:fkkernels}. We use the model F5 at redshift $z=0.38$.}
 	\label{fig:ScreeningsVsEFT}
 	\end{center}
 \end{figure}

\end{subsection}

\begin{subsection}{Screenings and EFT parameters} \label{sect:ScreeningVsEFT}

In cosmological large scale structure modeling, the screenings have played an important role and are necessary to properly match the simulated data \cite{Koyama:2009me,Aviles:2017aor}. However, these effects can be quite degenerate with EFT counterterms.  Using the F5 model at redshift $z=0.38$, in fig.~\ref{fig:ScreeningsVsEFT}  
we show with solid lines the ratio of the power spectrum multipoles $\ell= 0$ and 2 without screenings and considering full kernels. Here we have chosen typical values for the cosmological, EFT, bias and noise parameters. We notice that up to $k=0.2 \,h \text{Mpc}^{-1}$ the effect of the screenings is about $5\%$ in the monopole and almost a $10\%$ in the quadrupole. However, if we add a correction to the non-screened monopole of the form $P_0(k) \rightarrow P_0(k) + c_0 k^2 P_L(k)$ and one to the non-screened quadrupole $P_2(k) \rightarrow P_2(k) + c_2 k^2 P_L(k)$, the effect of the screenings is effectively counteracted. This is shown with dot-dashed lines, where we can see that the difference with the full kernels up to $k=0.2 \,h\text{Mpc}^{-1}$  is smaller than $0.5\%$, and for the quadrupole than the $2\%$. The oscillations in the dot-dashed lines appear because the added linear power spectrum is not IR-resummed, and they should disappear if this is properly done. 

This analysis suggests there is a high degeneracy between EFT counterterms and the chameleon screening in $f(R)$ that allows us to do not consider the latter in the analysis, which are very slow to compute. While we were finishing this work, an approximate method to treat the screenings accurately was proposed in Ref.~\cite{Euclid:2023bgs}. However, with the use of simulations we are capable of detecting the F5 signal with our simple prescription of ignoring the screenings. 

Typically, as larger are the MG strengths, the screening effects appear at lower $k$ values, which may be a signal of the breakdown of the fkPT approximation. For $f(R)$, this scale is given by $k_{M_1} = a \sqrt{M_1}$ \cite{Aviles:2018qotF}, where $M_1$ is given by eq.~\eqref{M1} and related to the effective mass of the new gravitational scalar degree of freedom by $m=\sqrt{M_1/3}$. Although it may be expected that the screenings operate mainly around this scale, limiting the extension of our formalism beyond it, we will not adhere to this \textit{rule-of-thumb} here. Instead, we will validate fkPT against simulations for the HS-$f(R)$ model.

Related to this topic, it's the concern of how much one can trust the  fkPT method for a specific theoretical model. Ideally, one should validate the theory against simulations, if available. A more economical option is to perform a comparison against the full perturbation theory results, as we have done in this section. However, perhaps the most efficient way to work is by using linear parameterizations, and expect that EFT counterterms absorb the effect of the screening non-linearities, as a practical renormalization of the EFT coefficients. This is plausible for power spectra of the form $P(k) = t(k)P^{\text{MG, non-screenings}}(k) + (1 - t(k))P^{\text{GR}}(k)$, where $t(k)$ is a transition function admitting an expansion $t(k) = 1 + ak^2 + \cdots$, such that the non-screened power spectrum, correctly modeled by fkPT, is obtained for large scales, GR is recovered for small scales, and intermediate scales having terms degenerate with EFT contributions.

\end{subsection}

\begin{subsection}{\text{fkpt} code}

\texttt{fkpt} is a C language code, public available at \href{https://github.com/alejandroaviles/fkpt}{https://github.com/alejandroaviles/fkpt}, that computes the one-loop redshift space tracers power spectrum using th fk-Perturbation Theory. It receives as input the $\Lambda$CDM linear power spectrum and the cosmological parameters at the desired output redshift $z$. It solves eq.~\eqref{Dplus} to obtain linear growth function $D_+(k,t)$ and growth factor $f(k,z)$. Then, it computes the MG power spectrum using   
\begin{equation} \label{PLtoPMG}
  P^\text{MG}(k,z) = \left( \frac{D_+^\text{MG}(k,z)}{D_+^\text{$\Lambda$CDM}(k,z)}\right)^2   P^\text{$\Lambda$CDM}(k,z).
\end{equation}
This is an excellent approximation for several MG models, as $f(R)$. However, the code can also receive directly the MG power spectrum obtained from another code. The code then splits the power spectrum in wiggle and non-wiggle pieces using the fast sine transform technique described in \cite{Hamann:2010pw}, then it computes the IR-resummed power spectrum given by eq.~\eqref{PsIR}. 

\texttt{fkpt} does not use an FFTLog method since we want flexibility that allows future addition of theories that do not reduce to $\Lambda$CDM at very large scales, and hence their kernels cannot be approximated as EdS when they are evaluated at small wave vectors. That is, our code treats the large scales exact, and also serves for computing the GR power spectrum using the exact $\Lambda$CDM kernels. Despite we use a brute force approach, our code takes about 0.5 seconds in a standard personal computer to compute a single power spectrum, hence being capable of explore the parameter space with MCMC in reasonable time.  

Other desired capabilities, as the Alcock-Paczyński effect and analytical marginalization, can be computed from the outside using a python interface that we provide together with the code in the github repository.


\end{subsection}

\end{section}

\begin{section}{Model validation} \label{sect:model_validation}

 \begin{figure}
 	\begin{center}
 	\includegraphics[width=6.2 in]{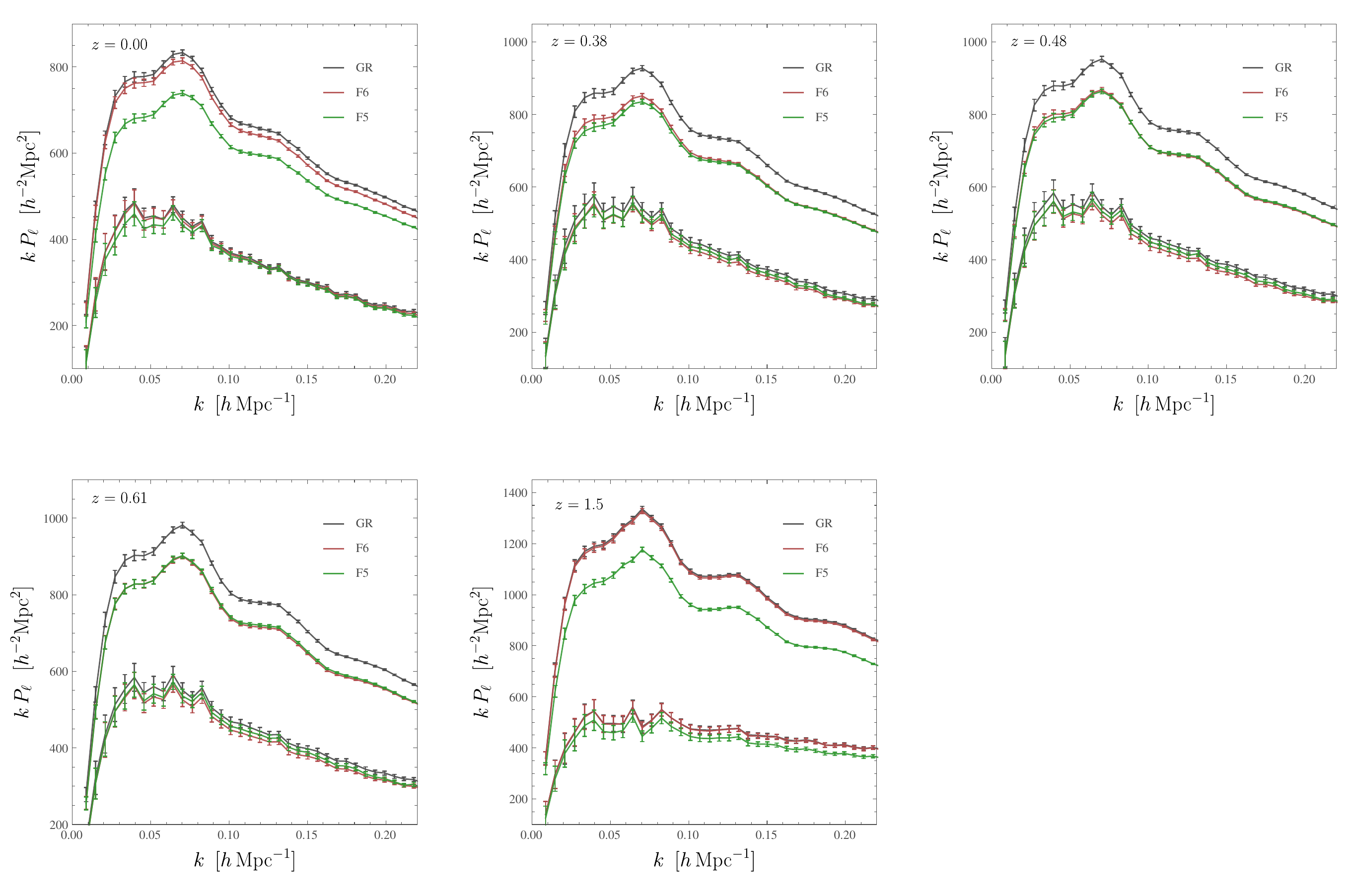}
 	\caption{MG-GLAM simulated halo power spectrum multipoles $\ell=0$ and 2. The central points are the mean of the 100 simulations and the error bars are their RMS which is further divided by $1/5$ to show the errors for the rescaled covariance \texttt{cov25}.  } 
 	\label{fig:multipoles}
 	\end{center}
 \end{figure}

Following the previous sections where we introduced the fk-Perturbation Theory framework and its implementation code \texttt{fkpt}, we are now ready to assess their performance by fitting simulated halo power spectra obtained from state-of-the-art $N$-body simulations. One of the main objectives of this study is to determine the capability of our method to successfully recover the MG signal in specific scenarios. We anticipate that our method is able to doing so when the signal is sufficiently strong, being effective only for F5 power spectra at low redshifts, or when we perform a joint analysis with tracers at different redshifts. Unfortunately, we were unable to detect the signal from F6, which is very close to GR.  On the other hand, our study also serves to gain insight on how much we can test GR with the full shape power spectrum, so we will devote some time to the analysis of $\Lambda$CDM simulations. 

 \begin{figure}
 	\begin{center}
 	\includegraphics[width=3 in]{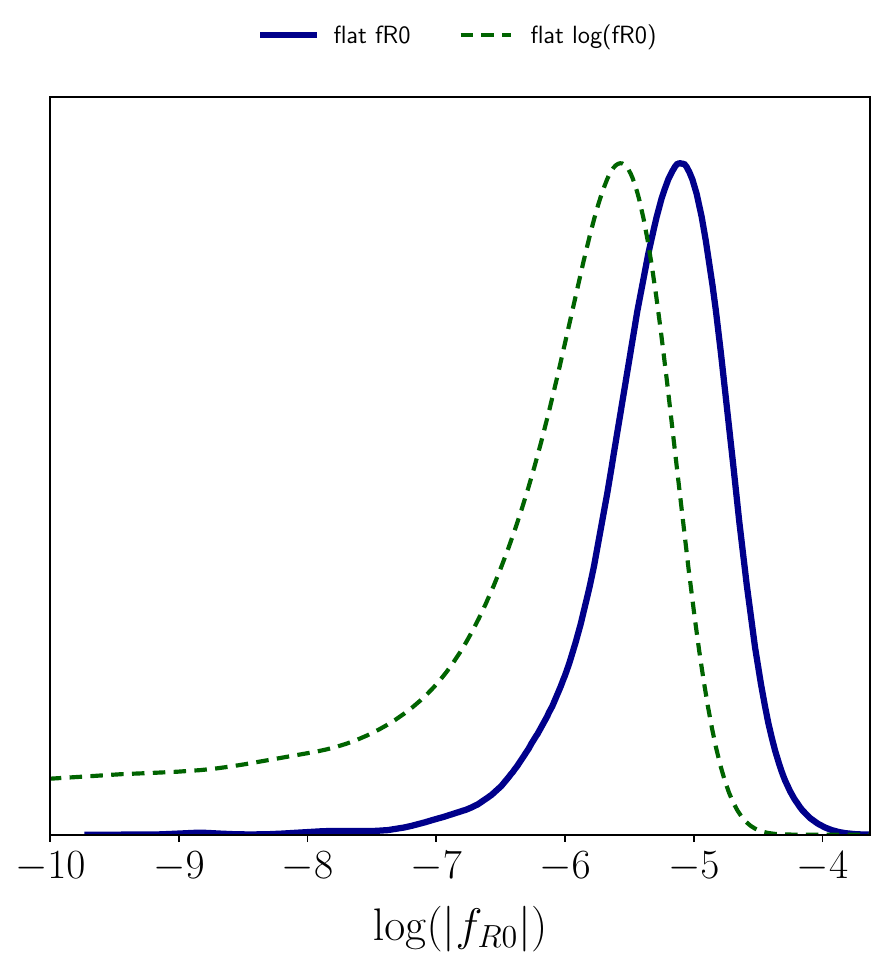}
 	\caption{\textit{Effect of priors on $f_{R0}$:} We show fittings to F5 model power spectrum  when opting for a flat prior on $f_{R0}$ over the interval (-0.01,0.01) (blue solid line) and a flat prior on $\log_{10}(|f_{R0}|)$ over the interval (-10,-1) (green dashed line). } 
 	\label{fig:prior}
 	\end{center}
 \end{figure}

Before discussing the results, we lay down a brief overview of the adopted $N$-body simulations. For these we use a suite of simulations performed with the code \textsc{MG-GLAM} \cite{Hernandez-Aguayo:2021kuh,Ruan:2021wup}, which is an extension of the parallel Particle-Mesh \textsc{GLAM} (GaLAxy Mocks) pipeline for fast generation of synthetic galaxy catalogs  \cite{Klypin:2017iwu}. The set of simulations contains runs with the following $\Lambda$CDM reference cosmology  
\[\{\Omega_b, \Omega_m, h, n_s, \ln(10^{10}A_s)\} = \{0.0486, 0.3089, 0.6774, 0.9667, 3.01887 \}\]
corresponding to the best fit values of Planck 2015 (last column in table 4 of \cite{Planck:2015fie}). 
Apart from the GR results, two instances of the HS-$f(R)$ model were simulated (among other MG models not considered in this work), corresponding to  $|f_{R0}|= 10^{-6}$ and $10^{-5}$, that we refer as to F6 and F5, respectively.  The simulations are performed over a cubic box of a comoving volume of $(1024 \,h^{-1}\,\text{Mpc} )^3$, with grid size of $4096^3$, and including $2048^3$ dark matter particles in each realization. All realizations were initialized with Zeldovich Approximation initial conditions at redshift $z=100$ using the $z=0$ power spectrum of the $\Lambda$CDM reference cosmology extrapolated to the initial redshift. This implies that the $\sigma_8$ value is different for the different models, but the primordial amplitude remains the same. These values are $\sigma_8^\text{GR}=0.8161$, $\sigma_8^\text{F6}=0.8292$ and  $\sigma_8^\text{F5}=0.8694$.
Gravitationally bound halos were identified using the bound density maxima (BDM) spherical overdensity halo finder \citep{2011ApJ...740..102K}, selecting halos within the mass range $10^{12.5} < M_h < 10^{13}\,h^{-1} \,M_\odot$.

For our discussions, we have selected five specific redshifts widely used in the literature. These include $z=0$ and $z=0.48$, which is utilized as a proxy of $z=0.5$. Furthermore, we considered redshifts $z=0.38$ and $0.61$ as they coincide with the distinct, non-overlapping bins known as $z_1$ and $z_3$ in the BOSS DR12 dataset \cite{BOSS:2016wmc}. Finally, $z=1.5$ corresponds to the redshift of QSOs in the BOSS DR16. These particular redshift values have been extensively employed in joint analyses of 2-point statistics.

In fig.~\ref{fig:multipoles} we show plots for the halo power spectrum at different redshifts with the error bars arising from the covariance matrix rescaled by a factor of $1/5$ as explained below. We notice 
there is not a clear pattern in the halo power spectra of different gravity theories. For the matter power spectrum, the MG models present more power than GR at all scales because the strength of gravity is larger in MG. However, for the halo spectrum the situation is very different because of the halo large scale bias. It is known from previous works  that bias evolution differs among MG models \cite{Hui:2007zh,Parfrey:2010uy,Lam:2012fa,Aviles:2018saf,Arnold:2018nmv}, with a tendency for smaller linear  bias $b_1$ as the MG strength increases, see for example fig.~4 of ref.~\cite{Aviles:2019fli}. This phenomenon further complicates the differentiation between MG models. Figure ~\ref{fig:multipoles} also emphasizes the significance of bias evolution in MG. Notably, for redshifts $z=0.38$, $0.5$ and $0.61$, models F5 and F6 are nearly indistinguishable, as they overlap within the error bars of our simulations, even after rescaling the covariance matrix. Conversely, at redshift $z=1.5$, models GR and F6 overlap, and model F5 apparently could be distinguished, however the observed differences are almost entirely due to a different large scale bias, since the clustering effects of MG are very small at such a large redshift. This can be confirmed by examining the ratio between the same multipole for two different models, e.g. GR to F5, which remains nearly constant with the wavenumber $k$. We conclude, that for our chosen halos the only redshift at which the three models could be clearly distinguished is $z=0$. 
This shows the importance of employing joint analysis of different redshifts or tracers when testing gravity.

\begin{center}
\begin{table*}
\ra{1.3}
\begin{center}
\begin{tabular} { l  c }

 Parameters &  Priors  \\
\hline
\vspace{0.15cm}

$\quad f_{R0}$ &  $\mathcal{U}(-0.01,0.01)$ \\ 
\vspace{0.15cm}

$\quad \omega_\text{cdm}$ & $\mathcal{U}(0.05, 0.2)$\\
\vspace{0.15cm}

$\quad \omega_\text{b}$ & $\quad \mathcal{G}(0.02230,0.00038) \, $\\
\vspace{0.15cm}

$\quad h$ & $\mathcal{U}(0.4,0.9)$   \\
\vspace{0.15cm}

$\quad \ln(10^{10}A_s)$ & $\mathcal{U}(2.0,4.0)$ \\
\vspace{0.15cm}

$\quad b_1$ & $\mathcal{U}(0.2, 3)$ \\
\vspace{0.15cm}

$\quad b_2$ & $\mathcal{U}(-10,10)$ \\
\vspace{0.15cm}

$\quad \alpha_0$ & $\mathcal{U}(-200,200)$ \\
\vspace{0.15cm}

$\quad \alpha_2$ & $\mathcal{U}(-200,200)$ \\
\vspace{0.15cm}

$\quad \alpha_0^\text{shot}$ & $\mathcal{U}(0,50)$ \\
\vspace{0.15cm}

$\quad \alpha_2^\text{shot}$ & $\mathcal{U}(-80,80)$ \\

\hline

\end{tabular}
\caption{Cosmological and nuisance parameters and Gaussian ($\mathcal{G}$) and uniform ($\mathcal{U}$) priors used for fitting the MG-GLAM simulated data. We fix the primordial spectral index to $n_s = 0.9667$. Tidal bias $b_{s^2}$ and $b_{3nl}$ are obtained from coevolution using eqs.~\eqref{coevbiases}.   }
\label{table:priors}
\end{center}
\end{table*}
\end{center}

For each of the examined data sets, we fit the mean of the 100 realizations. The power spectrum multipoles covariance of a single realization is determined by 
\begin{equation} \label{CovMat1}
    C_{\ell \ell'}(k_i,k_j) = \left\langle \Big(\bar{P}_{\ell}(k_i) - \hat{P}_{\ell}(k_i) \Big) \Big( \bar{P}_{\ell'}(k_j) - \hat{P}_{\ell'}(k_j) \Big)\right\rangle,
\end{equation}
where $\hat{P}(k_i)$ is the value of the power spectrum of a single realization at bin $k_i$ and $\bar{P}(k_i) = \langle \hat{P}(k_i) \rangle$  is the mean over the ensemble of realizations at bin $k_i$. However, since the volume of each simulation is $(1024 \, h^{-1}\,\text{Mpc})^3$, we rescale the covariance $C$ by factors of $1/25$ and $1/100$, considering effective volumes of $\sim 25  \,h^{-3}\text{Gpc}^3$ and   $\sim 100  \,h^{-3}\text{Gpc}^3$. We refer to these sets of data as \verb|cov25| and \verb|cov100|. With these we construct a Gaussian Likelihood $L \propto \exp(-\chi^2/2)$,  with 
\begin{equation}
   \chi^2 =  ( P^\text{theory} -  P^\text{sim})^\text{T} C_{N}^{-1}( P^\text{theory} -  P^\text{sim}) 
\end{equation}
with $P^\text{sim}=P_{\ell}^\text{sim}(k_i)$ the data vector, the $P^\text{theory}=P_{\ell}^\text{theory}(k_i)$ the model vector, and $C_{N}$ is the covariance matrix in eq.~\eqref{CovMat1},
\begin{equation}
  C_N =   \frac{1}{N} C_{\ell \ell'}(k_i,k_j),
\end{equation}
rescaled either by the factor $1/N$, with $N = 25$ or $100$.

Our fitting set consists of 11 parameters: four cosmological $\{h,\omega_c,\omega_b,\ln(10^{10}A_s)\}$, one MG parameter $f_{R0}$, two local biases $\{b_1,b_2\}$, two EFT counterterms $\{\alpha_0,\alpha_2\}$ and the two shot noise $\{\alpha_0^\text{shot}, \alpha_2^\text{shot} \}$. The spectral index $n_s$ is fixed to the simulation value. Meanwhile, tidal bias $b_{s^2}$ and third-order non-local bias $b_{3nl}$ are fixed by co-evolution theory \cite{Chan:2012jj,Baldauf:2012hs,Saito:2014qha} to
\begin{equation}\label{coevbiases}
 b_{s^2}= -\frac{4}{7}(b_1-1),\qquad b_{3nl}=\frac{32}{315}(b_1-1).
\end{equation}
These expressions are valid within GR when assuming local Lagrangian bias. However, previous works have found that these same relations yield good results for MG simulations \cite{Aviles:2020wme}, as well as in massive neutrino cosmologies \cite{Noriega:2022nhf}.

 \begin{figure}
 	\begin{center}
 	\includegraphics[width=4 in]{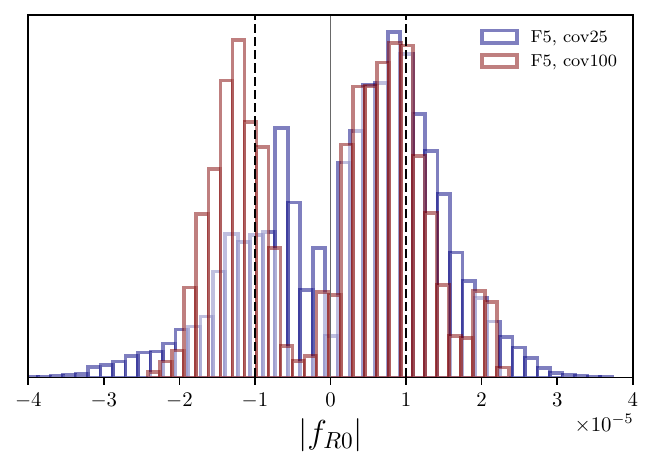}
 	\caption{F5 detection: We present the posterior distribution for fitting the F5 MG-GLAM simulations at redshift $z=0$ in the \texttt{cov25} and \texttt{cov100} cases. The parameter $f_{R0}$ is varied with a flat prior over the interval $(-0.01, 0.01)$. The confidence intervals for the absolute values are given in table \ref{table:z0cov25Cheng}. In our pipeline, the \texttt{fkpt} code treats only the absolute value of $f_{R0}$, ignoring its sign. This means that a signal detection would cause the MCMC chains to randomly sample the posterior around either the values $f_{R0}=-10^{-5}$ or $10^{-5}$ (shown as dashed vertical gray lines). For the sake of transparency, we do not smooth the MCMC in this plot. Our analysis provides unequivocal evidence that our pipeline successfully detects the MG F5 signal.} 
 	\label{fig:F5detection}
 	\end{center}
 \end{figure}

In our fittings we adopt uniform flat priors in all parameters with the exception of $\omega_b$, for which we use a Gaussian prior centered at the value of the simulations but a width given by Big Bang Nucleosynthesis (BBN) observations \cite{Aver:2015iza, Cooke:2017cwo}. None of the posterior distributions saturate the flat priors, indicating that they can be considered as uninformative. A list of all priors is provided in table \ref{table:priors}. We have chosen a flat prior on $|f_{R0}|$, instead of the perhaps more natural option of a flat prior over $\log|f_{R0}|$. We do this because our preliminary results have shown a better performance, getting a closer $ |f_{R0}|$ best fit value when fitting the F5 simulations for $z=0$, as shown in fig.~\ref{fig:prior}. In this plot we utilize a uniform prior $\mathcal{U}(-10,-1)$ for $\log|f_{R0}|$.

When reporting the parameters in figures and the main tables, we do it only for the cosmological parameters and the linear bias $b_1$. We omit $\omega_b$ for which the posteriors are entirely dominated by the prior, which is very tight. Furthermore, for the matter density we combine the baryons and cold dark matter abundances and use $\Omega_m$ instead of $\omega_m = \omega_b+\omega_c$ to avoid showing the trivial degeneracy with $h$.

 \begin{figure}
 	\begin{center}
 	\includegraphics[width=3.0 in]{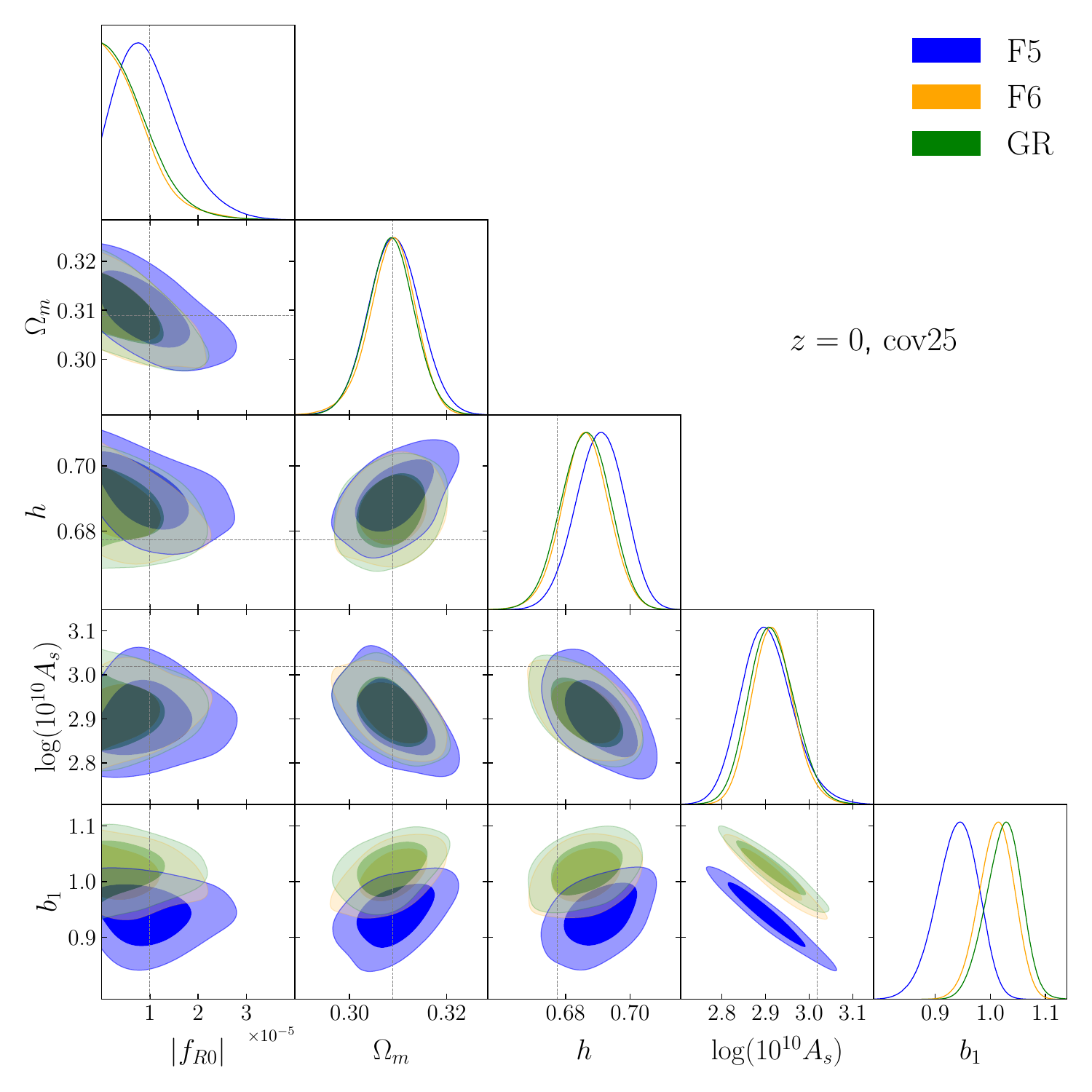}
 	\includegraphics[width=3.0 in]{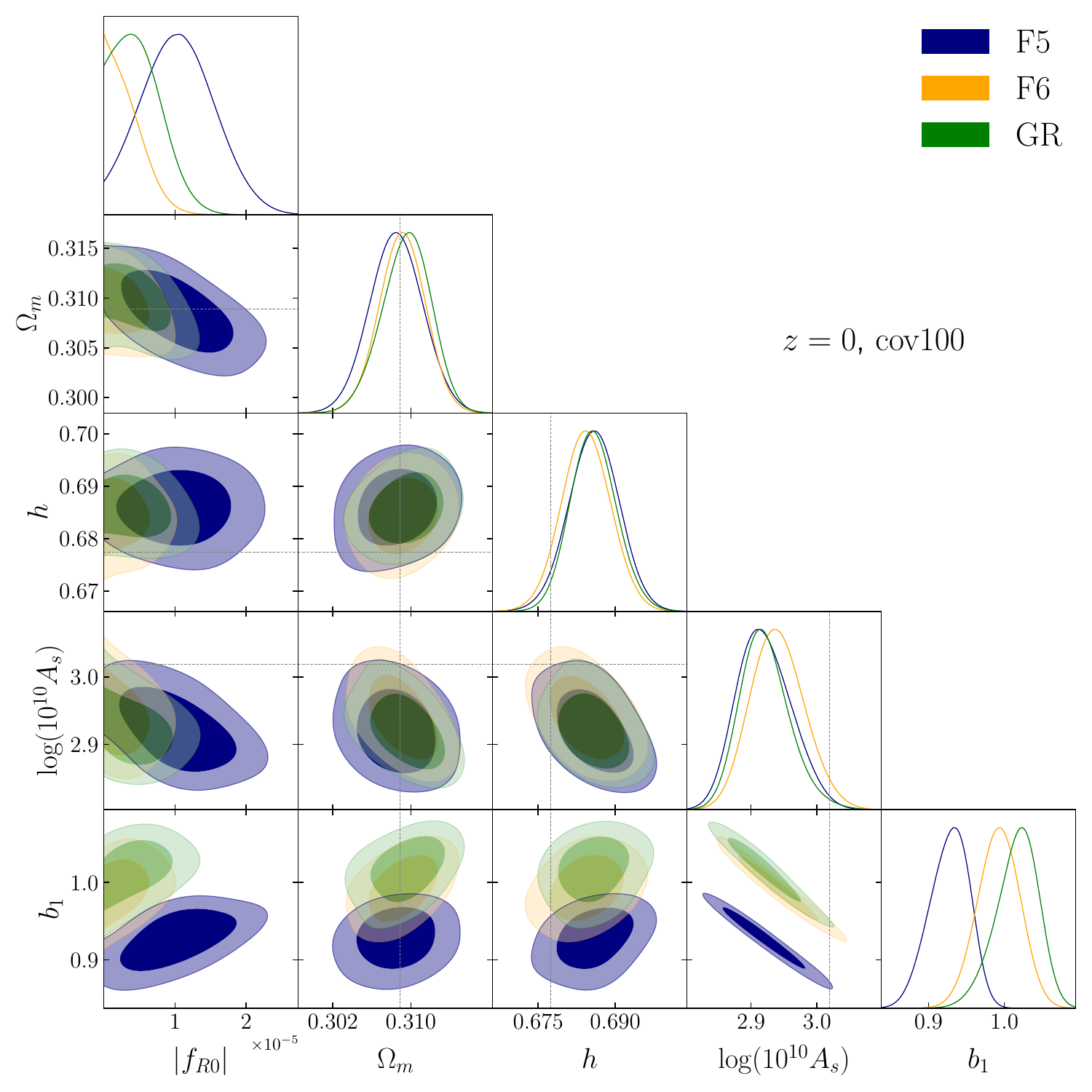}
 	\caption{Triangle plots for fitting GR, F6 and F5 simulations at redshift $z=0$, for the \texttt{cov25} case (left panel) and \texttt{cov100} (right panel). The shadows show the 0.68 and 0.95 confidence intervals. } 
 	\label{fig:z0cov25}
 	\end{center}
 \end{figure}

The linear $\Lambda$CDM power spectrum at redshift $z_0=0$ is obtained from the Einstein-Boltzmann code \texttt{CLASS}\footnote{\href{https://lesgourg.github.io/class_public/class.html}{https://lesgourg.github.io/class\_public/class.html}} \cite{Blas:2011rf} which serves as an input of \texttt{fkpt}, which first obtains the MG power spectrum using eq.\eqref{PLtoPMG}. Subsequently, the code computes the loop corrections in eqs.~\eqref{PsIR} and finally the halo power spectrum multipoles $\ell=0$ and $2$ through eq.~\eqref{Pells} that we compare against the simulated data.  To sample the parameter space, we perform MCMC runs with the code \texttt{emcee}\footnote{\href{https://emcee.readthedocs.io/}{https://emcee.readthedocs.io/}} \cite{ForemanMackey:2012ig} which is based in the affine-invariant ensemble sampler method \cite{2010CAMCS...5...65G}.  The contour and 1-dimensional posterior plots as well as the confidence intervals are computed using the \texttt{GetDist}  \texttt{Python} package \cite{Lewis:2019xzd}. For presentation purposes, in all figures with the exception of fig.~\ref{fig:F5detection} we use a \texttt{GetDist} smoothing scale of 0.7. However, the confidence intervals we present in all tables are computed without any previous smoothing.  

\begin{center}
\begin{table*}
\ra{1.7}
\begin{center}
\begin{tabular} { l  c  c  c}

$z=0$ &  F5 & F6  & GR \\
\hline

\underline{\texttt{cov25}}& $\qquad\qquad\qquad\qquad$& $\qquad\qquad\qquad\qquad$& $\qquad\qquad\qquad\qquad$\\

$|f_{R0}|$ & $1.03^{+0.38}_{-0.76}\,\times 10^{-5}$ &  $< 8.03\times 10^{-6}$  & $< 8.45\times 10^{-6}$\\

$\Omega_{m}$ & $0.3092\pm 0.0051 $   & $0.3088\pm 0.0046 $  &  $0.3085\pm 0.0047 $\\

$h$ & $0.6907\pm 0.0072 $   & $0.6861\pm 0.0070 $  &  $0.6862\pm 0.0073 $\\

$\ln(10^{10}A_s)$ & $2.903^{+0.054}_{-0.062}$   &  $2.916\pm 0.046  $ & $2.914^{+0.049}_{-0.055}   $ \\

$b_1$ & $0.940^{+0.041}_{-0.032}$   &  $1.011\pm 0.031$ & $1.024^{+0.034}_{-0.027}$ \\

\hline

\underline{\texttt{cov100}}& & & \\

$|f_{R0}|$ & $(1.06\pm 0.48) \times 10^{-5}$ & $< 4.21\times 10^{-6}$ & $ 5.1^{+2.1}_{-4.6}  \times 10^{-6}$   \\

$\Omega_{m}$ & $0.3085^{+0.0024}_{-0.0027}$  & $0.3091\pm 0.0023          $   & $0.3087^{+0.0046}_{-0.0029}$\\

$h$ & $0.6859\pm 0.0047$ & $0.6843\pm 0.0046$   &  $0.6855\pm 0.0043$ \\

$\ln(10^{10}A_s)$ & $2.920^{+0.031}_{-0.048} $ & $2.940^{+0.038}_{-0.043}$ &  $2.920^{+0.032}_{-0.039}   $\\

$b_1$ & $0.928^{+0.031}_{-0.018}$   &  $0.993\pm 0.027$ & $1.016^{+0.035}_{-0.018} $ \\

\hline

\bottomrule

\end{tabular}
\caption{One-dimensional 0.68 condifence intervals for fitting GR, F6 and F5 simulations at redshift $z=0$, for the cases \texttt{cov25} (top panel) and \texttt{cov100} (bottom panel). This table acompaines figure \ref{fig:z0cov25}.}
\label{table:z0cov25Cheng}
\end{center}
\end{table*}
\end{center}

In fig.~\ref{fig:F5detection} we show the posterior distribution when fitting the F5, MG-GLAM simulations at redshift $z=0$, considering both the \verb|cov25| case (blue lines) and \verb|cov100| case (red lines), and utilizing a maximum value $k_\text{max}=0.17 \,h\,\text{Mpc}^{-1}$ in the power spectrum. As detailed in Table \ref{table:priors}, we vary the MG parameter $f_{R0}$ with a flat prior ranging from $-0.01$ to $0.01$. However, our \texttt{fkpt} code only considers the absolute value of $f_{R0}$, being insensitive to its sign. We opt for this symmetric prior instead of a simpler range $(0, 0.01)$ to avoid encountering edge effects, which can arise when the MCMCs approach the boundaries of the interval. These boundary regions are precisely where we expect to find the signal, hence we aim to prevent any bias from such effects. We anticipate that a signal detection would cause the chains to randomly sample the posterior around either $f_{R0}=-10^{-5}$ or $10^{-5}$. This behavior is clearly observed in fig.~\ref{fig:F5detection}, particularly in the \verb|cov100| fitting case. Additionally, we omit to smooth the MCMC in this plot to avoid any visual ambiguity. Instead, we opt to exhibit a histogram. Overall, this analysis provides strong evidence that our theoretical model and numerical implementation have successfully detected the MG F5 signal.

In table \ref{table:z0cov25Cheng} we display the one-dimensional posterior confidence intervals when performing the analyses at redshift $z=0$ for the case of F5, as described above, as well as for F6 and GR. We notice that we recover all the cosmological parameters within the 1- or 2-$\sigma$ intervals, with the exception of $A_s$, for which we underestimate the true value.\footnote{We notice that this underestimation is very usual when comparing to BOSS data; 
see e.g. figure 6 in ref.~\cite{Ramirez:2023ads} for a comparison of the  estimated parameters using different full-modeling and other methods. This behavior may be attributed to the use of non optimal priors. However, this is not commonly observed with the use of simulated data. For that reason below we will use a different set of GR simulations for which we obtain a consistent primordial amplitude.}
Our table \ref{table:z0cov25Cheng} is accompanied by fig.~\ref{fig:z0cov25},  that shows triangular plots including the 2-dimensional contours and 1-dimensional distributions for the for \verb|cov25| case (left panel) and \verb|cov100| (right panel). We notice we obtain a 1-$\sigma$ detection of the MG-F5 signal for the case \verb|cov25|, and almost 2-$\sigma$ for \verb|cov100|. However, no detection is found in either case for the weaker gravity model F6. As explained above, in fig.~\ref{fig:multipoles} we displayed plots for the data fitted in this analysis, which suggest that the F6 and GR models yield very similar results on these halo mass range, while F5 can be clearly distinguished.

To compare our numerical fittings against a theoretical model, we calculated the linear large-scale halo bias using the Peak-Background Split (PBS) theory as described in \cite{Aviles:2018saf}.  
In scale-dependent MG, the critical threshold for collapse $\delta_c$---the minimum density fluctuation necessary for a space region to collapse and create a halo—varies as a function of the enclosed mass within that region. This differs from General Relativity (GR), where this threshold remains constant.
We compute $\delta_c(M)$ using the formulae of \cite{Kopp:2013lea}. Further, we use a Sheth-Tormen halo mass function, which has proven to be universal also for HS-$f(R)$ \cite{Aviles:2018saf} when seen as a function of the peak significance $\nu=\delta_c / \sigma_R$, instead of the variance $\sigma_R$, as is done in some works, e.g. \cite{Gupta:2021pdy}. For a redshift of $z=0$, we obtained the following values for the linear halo bias
\begin{equation}
b_1^\text{F5, PBS} = 0.92,\quad b_1^\text{F6, PBS} = 0.97, \quad b_1^\text{GR, PBS} = 1.01.    
\end{equation}
These values are in good agreement with our results as can be observed from table \ref{table:z0cov25Cheng}. 

 \begin{figure}
 	\begin{center}
 	\includegraphics[width=3.0 in]{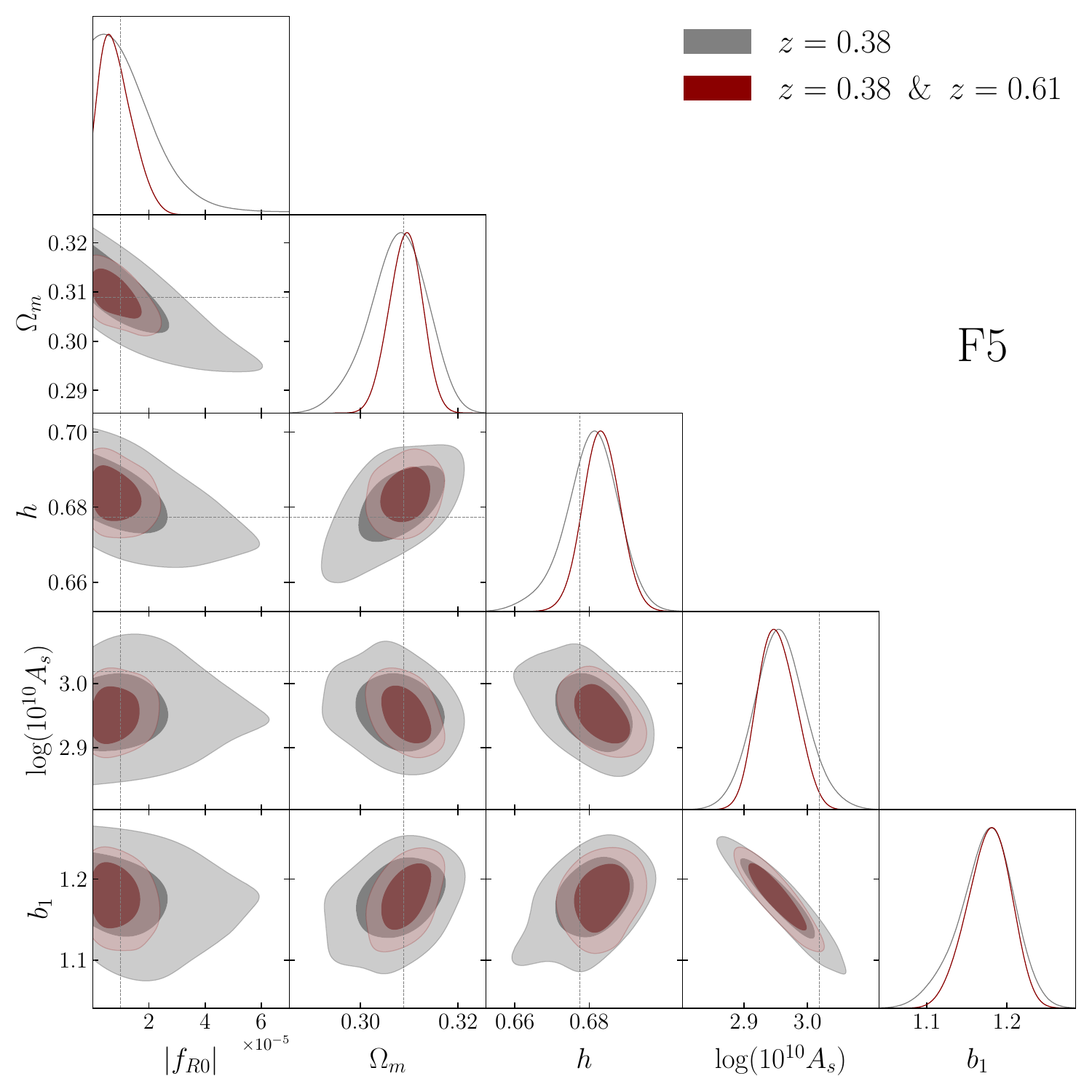}
 	\includegraphics[width=3.0 in]{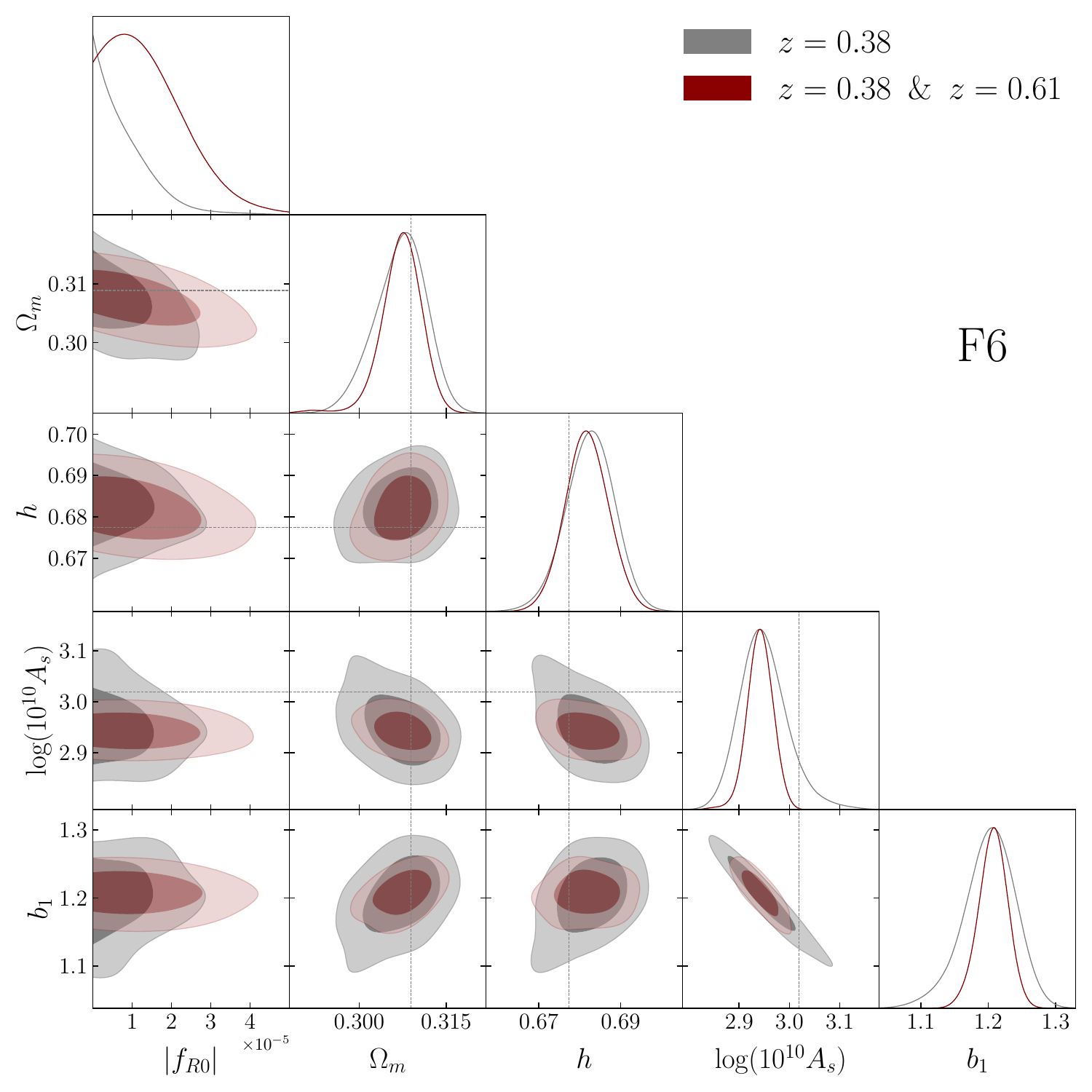}
 	\caption{Triangle contour plots from fits using F5 (left panel) and F6  (right panel) power spectra from simulated halo catalogs. The results are presented for redshift $z=0.38$ and for the joint analysis of redshifts $z=0.38$ and $z=0.61$. The covariance is rescaled as \texttt{cov25}. The shadows indicate the 0.68 and 0.95 confidence intervals. This figure is accompained by table \ref{table:redshifts}. } 
 	\label{fig:F5_zs}
 	\end{center}
 \end{figure}

\bigskip
To continue with the analysis, we performed fits to the F5 simulations at redshift 0.38 for covariance rescaled by a factor of 1/25 (\verb|cov25|). Unlike the redshift $z=0$, in this case we did not recover the MG signal, as can be seen in the left panel of fig.~\ref{fig:F5_zs} and in table \ref{table:redshifts}. For this reason, we  performed a joint analysis, named $z_{1,3}$, that includes halos at redshifts 0.38 and 0.61. In this case, we were able to obtain the value $|f_{R0}|=10^{-5}$ within the 0.68 confidence intervals. We have seen in Figure 2 and discussed above that it is difficult, if not impossible, to discern between F6 and F5 models for these two redshifts. Therefore, we also ran chains for F6 and show our results in the right panel of fig.~\ref{fig:F5_zs} and in table \ref{table:redshifts}. We notice that this analysis also detects a MG signal, but it is also located close to $|f_{R0}|=10^{-5}$, as we found for F5. However, the estimated linear bias is indeed different for cases F5 and F6, and these are in agreement with our theoretical results obtained using PBS, which are 
\begin{equation}
    b_1^\text{F5, PBS} = 1.14\quad \text{and} \quad b_1^\text{F6, PBS} = 1.19.
\end{equation}

\begin{center}
\begin{table*}
\small
\ra{1.7}
\begin{center}
\begin{tabular} { l  c c c c }

 Parameter &  F5, $z_1$ &   F5, $z_{1,3}$ &  F6, $z_1$ &   F6, $z_{1,3}$\\
\hline
$f_{R0}$       &  $< 1.67\cdot 10^{-5}$  &  $0.80^{+0.31}_{-0.75}\,\times 10^{-5}$ & $< 1.03\cdot 10^{-5}$ &  $0.83^{+0.27}_{-0.72}\,\times 10^{-5}$\\

$\Omega_{m}$   &  $0.3077^{+0.0064}_{-0.0050}$ & $0.3093^{+0.0035}_{-0.0031}$ & $0.3072^{+0.0049}_{-0.0036}$ & $0.3081^{+0.0027}_{-0.0024}$\\

$h$            & $0.6806^{+0.0073}_{-0.0059}$  &  $0.6833\pm 0.0048          $ & $0.6826^{+0.0061}_{-0.0053}$& $0.6823^{+0.0039}_{-0.0050}$\\

$\ln(10^{10}A_s) \qquad$ & $2.957\pm 0.040            $ & $2.952^{+0.026}_{-0.034}   $& $2.949^{+0.035}_{-0.051}$ & $2.941\pm 0.024            $\\

$b_1$             & $1.174^{+0.035}_{-0.027}$ & $1.178^{+0.030}_{-0.023}$ & $1.202^{+0.040}_{-0.031}$ &  $1.210^{+0.023}_{-0.020}   $\\
\hline

\bottomrule

\end{tabular}
\caption{One-dimensional constraints in fits using F5 and F6 simulations. The results are presented for redshift $z_1=0.38$ and for the joint analysis $z_{1,3}$, the later corresponding to both redshifts $z=0.38$ and $z=0.61$. The covariance is rescaled as \texttt{cov25}. The error bars indicate the 0.68 confidence intervals. The reported $b_1$ for the $z_{1,3}$ case represents the linear bias of the halos at $z=0.38$. This table accompanies figure \ref{fig:F5_zs}. }
\label{table:redshifts}
\end{center}
\end{table*}
\end{center}

\begin{center}
\begin{table*}
\small
\ra{1.7}
\begin{center}
\begin{tabular} { l  c c c c}

  &  $z_0$, free $f_{R0}$ & $z_0$, $f_{R0}=0$  & $z_{1,3}$, free $f_{R0}$ & $z_{1,3}$, $f_{R0}=0$  \\
\hline
$f_{R0}$        & $< 8.74\times 10^{-6}$   &  ---  & $< 7.04\times 10^{-6} $ & ---\\

$\Omega_{m}$    & $0.3085\pm 0.0046$   &  $0.3113\pm 0.0038 $   & $0.3086^{+0.0031}_{-0.0025}$ & $0.3093\pm 0.0026 $\\

$h$             & $0.6860\pm 0.0072$   &   $0.6871^{+0.0065}_{-0.0074}$  & $0.6822\pm 0.0051$ &  $0.6825\pm 0.0050 $\\

$\ln(10^{10}A_s)$ & $2.915^{+0.042}_{-0.056}   $   &  $2.917^{+0.040}_{-0.058}$  & $2.949^{+0.023}_{-0.029}$ & $2.956^{+0.023}_{-0.035}$\\

$b_1$   & $1.023^{+0.035}_{-0.027}   $   & $1.018^{+0.041}_{-0.026}   $   &  $1.266^{+0.026}_{-0.023}$ & $1.259^{+0.033}_{-0.022}$\\

\hline

\bottomrule

\end{tabular}
\caption{\textit{GR:} One-dimensional constraints on the parameter $f_{R0}$ in fits using GR simulations. The results are presented for redshift $z_0=0$ and for the joint analysis $z_{1,3}$, the later corresponding to redshifts $z=0.38$ and $z=0.61$. The covariance is rescaled as \texttt{cov25}. The error bars indicate the 0.68 confidence intervals. The reported $b_1$ for the $z_{1,3}$ case represents the linear bias of the halos at $z=0.38$.}
\label{table:GR}
\end{center}
\end{table*}
\end{center}

 \begin{figure}
 	\begin{center}
 	\includegraphics[width=3.0 in]{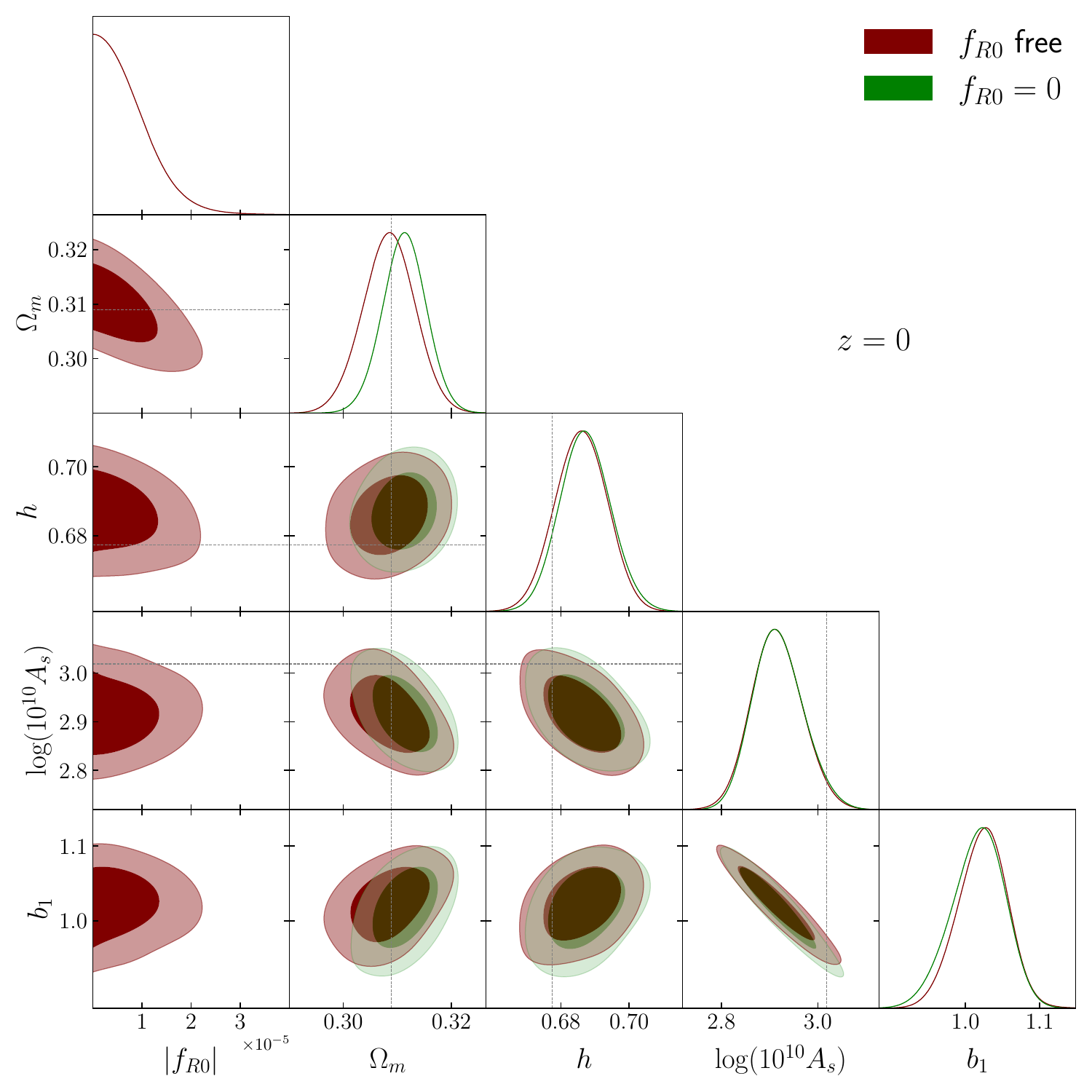}
 	\includegraphics[width=3.0 in]{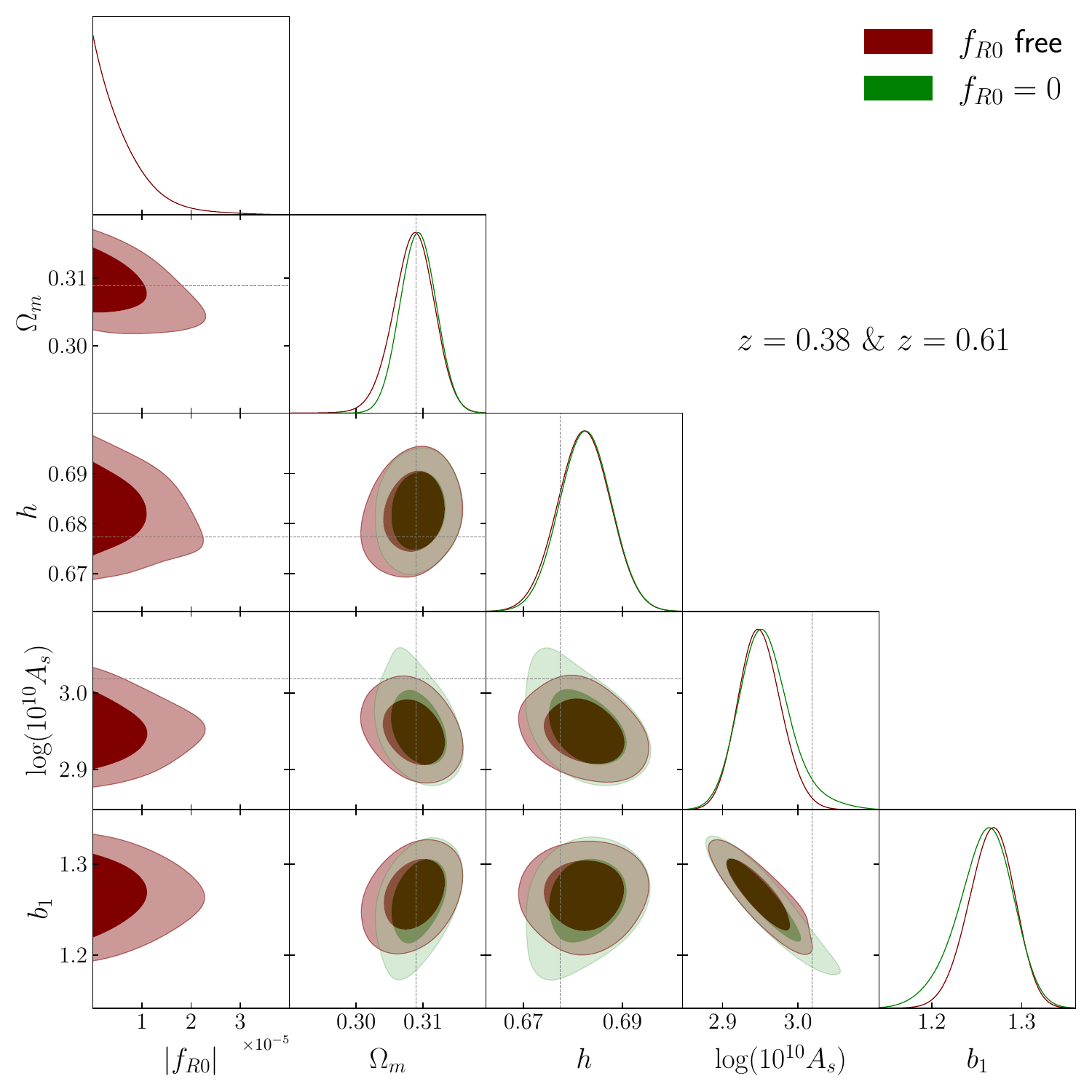}
 	\caption{Triangle contour plots from fits using GR for redshift $z=0$ (left panel) and joint analysis including redshifts $z=0.38$ and $0.61$ (right panel). We show the results when varying and kept fixed $f_{R0}$. The covariance matrix is rescaled as \texttt{cov25}. The shadows indicate the 0.68 and 0.95 confidence intervals. This figure accompanies  table \ref{table:GR}. } 
 	\label{fig:GRcomparisons}
 	\end{center}
 \end{figure}


Our approach can effectively estimate parameters in a standard $\Lambda$CDM cosmology, similar to other codes such as \texttt{Class-PT}, \texttt{PyBird}, \texttt{FOLPS}$\nu$, and \texttt{Velocileptors}. However, our method offers a potential advantage for future surveys by employing the exact $\Lambda$CDM kernels within our \texttt{fkpt} code, in contrast to the commonly used EdS kernels when fitting data. To demonstrate this, we conducted an analysis fitting General Relativity (GR) simulations while deactivating the effects of MG. This was achieved by setting $f_{R0}$ to a very small value, such as $10^{-10}$ or any other very small number in \texttt{fkpt}. We performed this analysis for two scenarios: the redshift $z_0=0$ and a joint analysis of two redshifts, $z_{1,3}$, including redshifts 0.38 and 0.61. The outcomes are presented in table \ref{table:GR} and fig.~\ref{fig:GRcomparisons}. We successfully recovered the cosmological parameters within the 2-$\sigma$ confidence intervals, with the exception of the underestimated $A_s$. To explore the effects of activating MG, we repeated the fittings with a free parameter $f_{R0}$. The results, shown also in fig.~\ref{fig:GRcomparisons} and Table \ref{table:GR}, are somewhat surprising. We observe no significant differences in the posterior distributions, except for a slight offset in the best fit of $\Omega_m$ and broadening in its distribution. 
This indicates a lack of substantial degeneracy between the cosmological parameters and $f_{R0}$ at large scales, with the possible exception of $\Omega_m$. 

Finally, we want to compare the dependence of our fittings on the maximal wave-number. The results are presented in fig.~\ref{fig:kmaxs}, where we show only the means of the full-shape analyses with $k_\text{max}=0.15$, 0.16, 0.17, 0.18, 0.19, 0.20 and $0.21 \,h\,\text{Mpc}^{-1}$ an their 1-$\sigma$ error bars. For these analyses we have used the F5 model simulations at $z=0$ with the maximum available volume given by the cov100 rescaling. This plot further justify our baseline choice of $k_\text{max}=0.17 \,h\,\text{Mpc}^{-1}$.

 \begin{figure}
 	\begin{center}
 	\includegraphics[width=4.0 in]{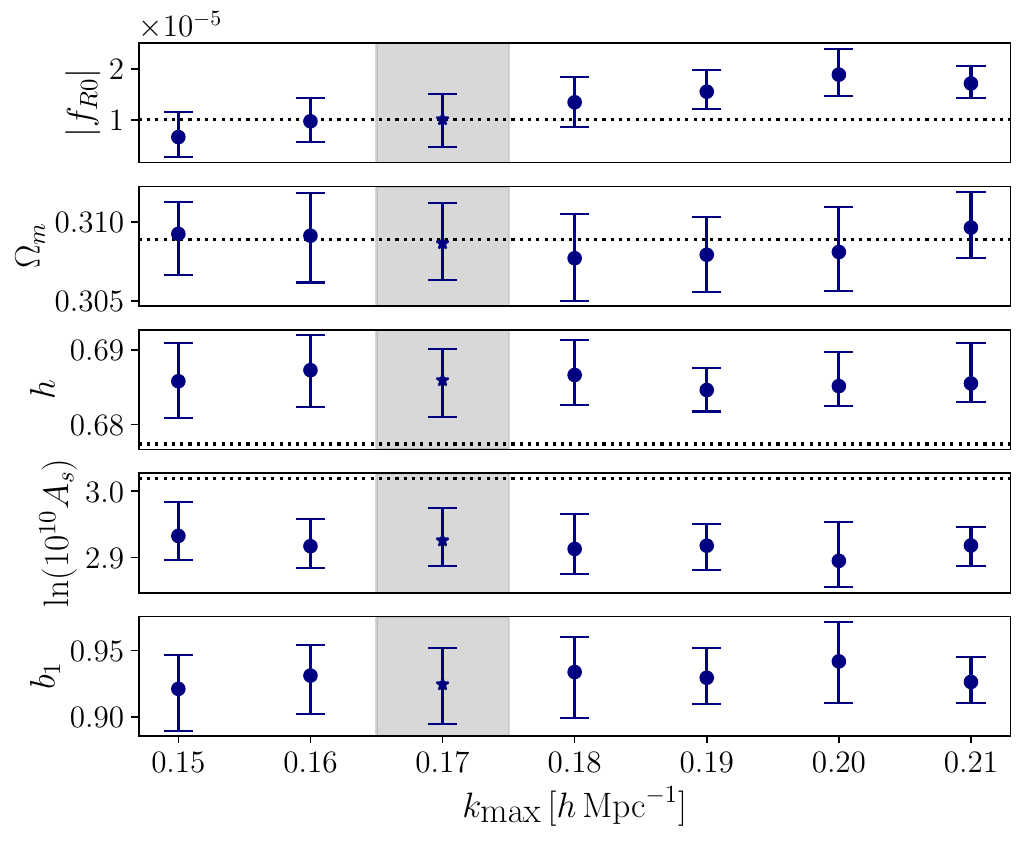}
 	\caption{Impact of  $k_\text{max}$ in the fkPT full-shape analysis, when fitting the \textsc{MG-GLAM} simulations, with F5 and redshift $z=0$. The dots indicate the means and the bars are the 1-$\sigma$ errors. The shadowed region corresponds to our baseline choice of $k_\text{max}=0.17 \,h\,\text{Mpc}^{-1}$ throughout this work.} 
 	\label{fig:kmaxs}
 	\end{center}
 \end{figure}

\subsection{Comparing to \textsc{NSeries} simulations}

Finally, since we are worried about the lack of precision in our fittings to \textsc{MG-GLAM} simulations, particularly in $A_s$ and to a lower extent into $h$, we opt to use a larger set of simulations, although they exist only for GR. 
Specifically, in this subsection we utilize the cubic boxes of the \textsc{NSeries} galaxy mocks, comprising 7 realizations, each one with a volume of $V_1=(2.7 \,h^{-1}\text{Gpc})^3$ \cite{BOSS:2016wmc}.\footnote{Available at \href{https://www.ub.edu/bispectrum/page12.html}{https://www.ub.edu/bispectrum/page12.html}.} These simulations were initially generated to investigate systematic effects within the BOSS data pipeline. The expanded volume offered by this dataset grants us greater confidence in testing our perturbative theoretical model, which extracts the cosmological information from the larger scales where simulations with smaller sizes, as \textsc{MG-GLAM}, are not optimal. The cosmological parameters are 
$\Omega_M = 0.286$,  $n_s = 0.97$, $h = 0.7$, $\ln(10^{10} A_s) = 3.06619$ and $\Omega_b=0.047$. The covariance matrix is constructed using the NGC 1,000 \textsc{EZmocks} catalogues \cite{Chuang:2014vfa} available at the same URL.   

We fit to the mean galaxy power spectra of the 7 realizations, with the covariance rescaled to the maximum allowed volume ($N=7$), such that the effective volume of the simulations is $V_7=137.8\,(h^{-1}\text{Gpc})^3$. This analysis besides  permitting us to have better confidence at the large scale fittings, allow us to test our modeling in a limiting case of unrealistic small errors. In fig.~\ref{fig:GRcomparisonsNseries} we show triangular plots for our fittings when letting free $|f_{R0}|$ and when keep it fixed to zero, these are accompanied by table \ref{table:GR_Nseries}. We notice that our fits are quite precise, matching the parameter of the simulations with high accuracy and precision even when the effects of MG are allowed.

 \begin{figure}
 	\begin{center}
 	\includegraphics[width=3.0 in]{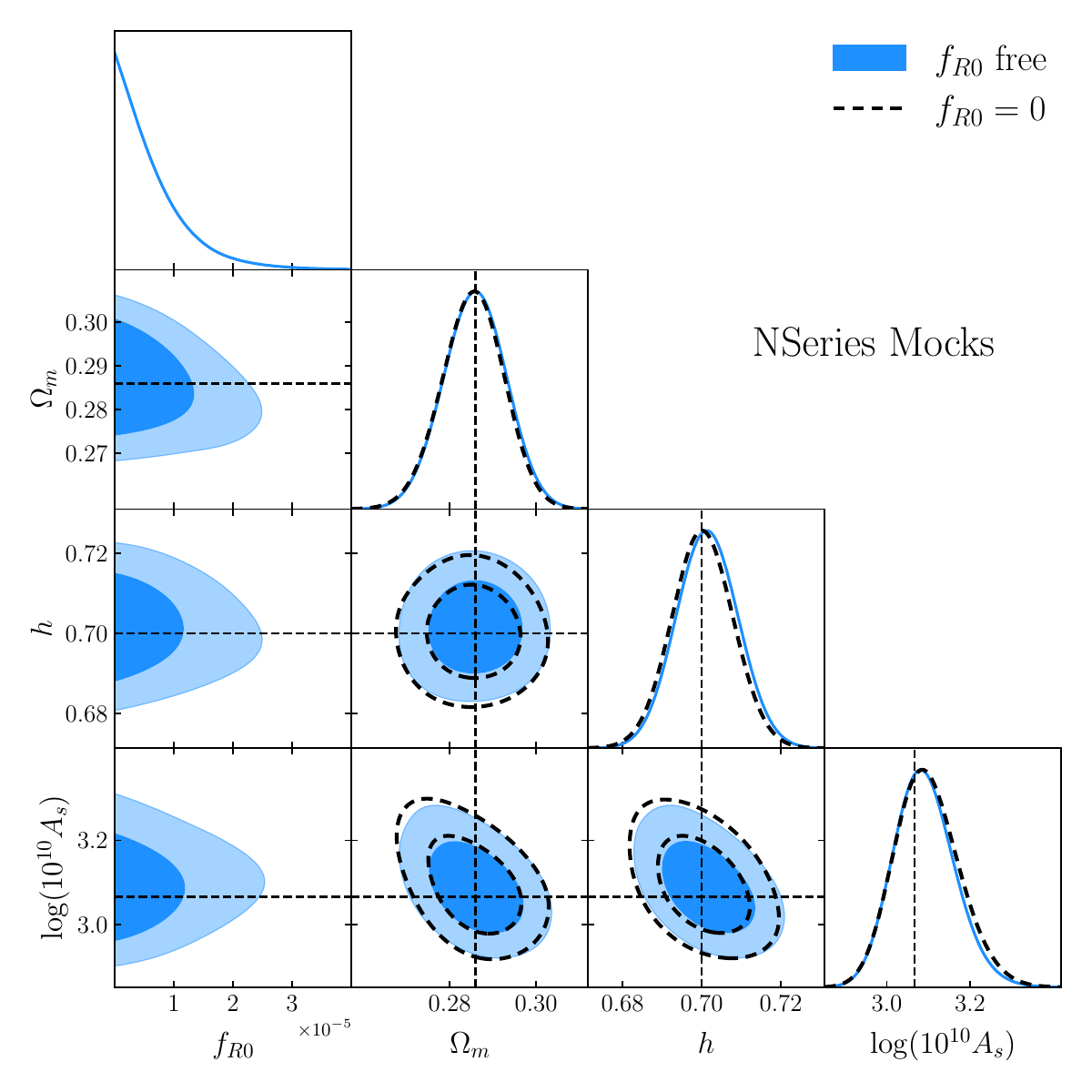}
 	\caption{Triangle contour plots from fits using \textsc{NSeries} GR simulations for redshift $z=0.5$.  This figure accompanies  table \ref{table:GR_Nseries}. } 
 	\label{fig:GRcomparisonsNseries}
 	\end{center}
 \end{figure}

\begin{center}
\begin{table*}
\small
\ra{1.7}
\begin{center}
\begin{tabular} { l  c c }

  & $\quad$ \textsc{NSeries}, free $f_{R0}$ $\quad$ & $\quad$ \textsc{NSeries}, $f_{R0}=0$ $\quad$ \\
\hline
$|f_{R0}|$        & $< 7.28\cdot 10^{-6}       $   &  ---  \\

$\Omega_{m}$    & $0.2858\pm 0.0069          $   &  $0.2854\pm 0.0068          $  \\

$h$             & $0.7014\pm 0.0074          $   &   $0.7002\pm 0.0074          $  \\

$\ln(10^{10}A_s)$ & $3.089^{+0.067}_{-0.077}   $   &  $3.095^{+0.071}_{-0.082}   $  \\

$b_1$   & $1.877^{+0.089}_{-0.080}   $   & $1.864^{+0.095}_{-0.084}   $   \\

\hline

\bottomrule

\end{tabular}
\caption{\textit{NSeries:} One-dimensional constraints on the parameter $f_{R0}$ in fits using \textsc{NSeries} GR simulations. The results are presented for redshift $z=0.5$. We fit the mean power spectrum of the 7 cubic boxes each with volume $V_1=(2.7 \,h^{-1}\text{Gpc})^3$.}
\label{table:GR_Nseries}
\end{center}
\end{table*}
\end{center}

\end{section}

\begin{section}{Conclusions} \label{sect:conclu}

In recent years, full-shape methods, also known as full-modeling or direct-fit, have become a standard approach for extracting cosmological information from galaxy surveys. This shift from the fixed-template classical-analysis resulted from advancements in the EFT of Structure Formation. This framework extends PT by incorporating biases, counterterms and shot noise parameters, as well as the use of IR-resummations to model the smearing of the BAO. Much of this progress has assumed a $\Lambda$CDM model where linear growth is scale-independent. While this approach adequately describes scenarios involving additional scales which produce small effects in clustering, such as those with massive standard model neutrinos, it may fall short for more generalized MG theories. In these theories, the two scalar gravitational potentials (in Newtonian Gauge) differ even in the absence of anisotropic stresses in the matter content, introducing scale-dependent terms into the Poisson equation. The pioneering work by \cite{Koyama:2009me} introduced perturbative methods to accommodate these new scales within the kernels, further developed in various scenarios in the literature.

However, a major challenge of these methods lies in the absence of algebraic expressions for the perturbative kernels. Instead, they must be derived by solving differential equations for each wave-vector configuration and each set of cosmological parameters. This significantly slows down the computation of statistic one-loop corrections, making parameter exploration, such as through MCMC, exceedingly time-consuming. The present work addresses this issue building-upon the formalism introduced in \cite{Aviles:2021que,Noriega:2022nhf} for massive neutrinos. We identify two types of contributions within the PT kernels that are absent in the EdS model. Firstly, the introduction of scale and time-dependent functions $\mathcal{A}$ and $\mathcal{B}$, and their third order counter-parts, due to the deviation of the growth factor $f$ from being equal to $\Omega^{1/2}(a)$ and because of the screening contributions that should drive the theory to GR at high $k$. While these functions exist in the $\Lambda$CDM model, they only depend on time in that context. Secondly, contributions arise from the scale-dependence of the growth factors $f=f(k,t)$, resulting in differences between velocity and density fields at linear order, encapsulated in $\theta^{(1)}(\vk,t) = (f(k,t)/f(k=0,t)) \delta^{(1)}(\vk,t)$. Due to the advection of large scale density fields, this property is inherited to higher orders in the perturbative kernels.

In scenarios such as  HS-$f(R)$ models, the first type of contribution is predominantly influenced by non-linear screening effects, often degenerate with EFT counterterms. Thus, we approximate these functions with their largest scale values, computed where all wave-vector arguments go to zero. Notably, for scale-dependent theories converging to the $\Lambda$CDM model at large scales, the approximated values correspond to the genuine $\Lambda$CDM values, unlike the EdS approximation where these functions equal unity. Hence, this method also serves us to test GR models using exact kernels.

Our approach, named $f(k)$-Perturbation Theory (fkPT), keeps the factors $f(k)/f(k=0)$ that represent the dominant contribution to the kernels, while maintaining the rest of the features as in their large scale limit. We have developed and released the \texttt{fkpt} code, enabling the computation of the redshift space power spectrum for scale-dependent MG.
We validate our method and code using the \textsc{MG-GLAM} simulations for HS-$f(R)$ models F6 and F5, alongside GR. While we successfully recovered the MG signal for F5 at redshift $z=0$, indicating its significant impact on the power spectrum, we were not able of doing so at $z=0.5$, since the MG signal is weaker in this case. However, we obtain the signal at higher redshifts through a joint analysis of $z_1=0.38$ and $z_3=0.61$. Additional examinations using DESI-like simulations and estimation of parameters using real data will be addressed in a future work.

Additionally, the evolution of large-scale bias differs across various gravity theories, suggesting different values of $b_1$ for the halos used in our analysis. We utilized the Peak-Background-Split formalism with a Sheth-Tormen mass function to theoretically derive large scale bias by using non-constant threshold density $\delta_c(M)$. Our analytical results coincide reasonably well with values obtained through a MCMC analysis, further demonstrating the efficacy of our method in recovering the MG signal from simulated data.

However, we acknowledge a limitation in our ability to recover the amplitude of the primordial perturbations. This shortcoming may be attributed to the modest size ($L=1024 \, h^{-1} \,\text{Gpc}$) of the \textsc{MG-GLAM} simulations, insufficient for thoroughly testing the large scales that are crucial for extracting cosmological information from full-shape analysis. Further, to find the MG signal we were forced to rescale the covariance by factors of $1/25$ and $1/100$, reaching effective volumes up to $V=100 \, (h^{-1}\,\text{Gpc})^3$, for which the simulations may not be sufficiently accurate. Therefore, we additionally use the cubic boxes of the \textsc{NSeries} simulations, primarily used to validate the BOSS pipelines. Despite being available only for GR, we successfully utilized these simulations, allowing the MG parameter $|f_{R0}|$ to vary freely and achieving excellent results using our methodology.

The advent of the Dark Energy Spectroscopic Instrument (DESI) and Euclid spectroscopic surveys opens new opportunities to test models beyond the standard $\Lambda$CDM with the clustering of galaxies. In particular, MG models that influence the late times clustering can elude probes based on CMB observations, or even kinematics tests in cosmological distances imposed by, e.g, Supernovae type Ia. Hence, new non-linear methodologies that account for additional scales, but at the same time being sufficiently fast to be implemented in MCMC samplers for parameter estimation, would prove valuable for investigating gravitational effects in the forthcoming years. Future endeavors in this line of research will steer towards this objective, together with the joint utilization of diverse cosmological probes.

\end{section}

\acknowledgments

AA and HN are supported by Ciencia de Frontera grant No. 319359.
AA, JLCC, and MARM~acknowledge support by CONACyT project 283151. 
BL acknowledges support by the UK STFC Consolidated Grants ST/P000541/1, ST/T000244/1, ST/X001075/1. 

The MG-GLAM simulations and MCMC analysis in this work used the DiRAC@Durham facility managed by the Institute for Computational Cosmology on behalf of the STFC DiRAC HPC Facility (www.dirac.ac.uk). The equipment was funded by BEIS capital funding via STFC capital grants ST/K00042X/1, ST/P002293/1, ST/R002371/1 and ST/S002502/1, Durham University and STFC operations grant ST/R000832/1. DiRAC is part of the National e-Infrastructure.

\appendix

\begin{section}{Third order kernels}\label{app:3rdOrder}

The transverse part of the third order Lagrangian displacement kernel is given by
 \begin{align} \label{LPTK3order}
  L^{(3)}_i(\vk_1,\vk_2,\vk_3) &=  \frac{k^i}{k^2}
    \Bigg\{ \frac{5}{7} \left( \mathcal{A}^{(3)} -\mB^{(3)}
              \frac{(\vk_2 \cdot \vk_3)^2}{k^2_2 k^3_2} \right) 
             \left( 1-  \frac{(\vk_1 \cdot \vk_{23})^2}{k_1^2 k_{23}^2} \right)  \nonumber\\*
    &  \quad -\frac{1}{3} \left( \, \mathcal{C}^{(3)}  -3 \mathcal{D}^{(3)} \frac{(\vk_2 \cdot \vk_3)^2}{k^2_2 k^2_3} 
     + 2 \mathcal{E}^{(3)} \frac{(\vk_1 \cdot \vk_2)(\vk_2 \cdot \vk_3)(\vk_3 \cdot \vk_1)}{k_1^2 k^2_2 k^2_3} \, \right)
     \Bigg\},                             
 \end{align} 
 with the scale- and time-dependent functions given by
\begin{align}
 \mathcal{A}^{(3)},\mathcal{B}^{(3)}(\vk_1,\vk_2,\vk_3) 
     &= \frac{7}{5}  \frac{D^{(3)}_{\mathcal{A},\mathcal{B}}(\vk_1,\vk_2,\vk_3)}{D_{+}(k_1)D_{+}(k_2)D_{+}(k_3)}, \\
 \mathcal{C}^{(3)},\mathcal{D}^{(3)},\mathcal{E}^{(3)}(\vk_1,\vk_2,\vk_3) 
     &= \frac{  D^{(3)}_{\mathcal{C},\mathcal{D},\mathcal{E}}(\vk_1,\vk_2,\vk_3)}{D_{+}(k_1)D_{+}(k_2)D_{+}(k_3)},
\end{align} 
and third order growth functions
\begin{align} 
 \big(\T - A(k)\big)D^{(3)}_\mathcal{A} &=   3 D_+(k_1) \big(A(k_1) + \T - A(k)\big)D^{(2)}_\mA(\vk_2,\vk_3)  , \\
 \big(\T - A(k)\big)D^{(3)}_\mathcal{B} &=    3 D_+(k_1) \big(A(k_1) + \T - A(k)\big)D^{(2)}_\mB(\vk_2,\vk_3)  , \\
 \big(\T - A(k)\big)D^{(3)}_\mathcal{C} &=    9 D_+(k_1) \big(A(k_1) + \T - 2 A(k)\big)D^{(2)}_\mA(\vk_2,\vk_3) \nonumber\\
                              &\quad   -3 A(k) D_+(k_1)D_+(k_2)D_+(k_3)  \nonumber\\
                              &\quad  + 3 K^{(3)}_\text{FL}(\vk_1,\vk_2,\vk_3) D_+(k_1)D_+(k_2)D_+(k_3), \\                          
 \big(\T - A(k)\big)D^{(3)}_\mathcal{D} &=    3 D_+(k_1) \big(A(k_1) + \T - 2 A(k)\big)D^{(2)}_\mB(\vk_2,\vk_3) \nonumber\\
                              &\quad     +3 A(k) D_+(k_1)D_+(k_2)D_+(k_3)  , \\
 \big(\T - A(k)\big)D^{(3)}_\mathcal{E} &=    3 \big(3 A(k_1) - A(k)\big)D_+(k_1)D_+(k_2)D_+(k_3),
\end{align}  
where the third order FL kernel is given by
\begin{align} \label{FL3} 
 &K^{(3)}_\text{FL}(\vk_1,\vk_2,\vk_3) = 3 (A(k)-A(k_1)) \left[ k_{23}^iL^{(2)}_i(\vk_2,\vk_3) \frac{\vk_1 \cdot \vk_{23}}{k_{23}^2}   
 -2 \frac{(\vk_1\cdot \vk_2)(\vk_1\cdot \vk_3)}{k_2^2 k_3^2} \right]\nonumber\\
 &\quad + 3 \big( A(k) - A(k_{23}) \big) \frac{\vk_{1}\cdot\vk_{23}}{k_1^2} 
      \left[  1 + 2\frac{(\vk_2\cdot \vk_3)}{k_3^2} + \frac{(\vk_2\cdot \vk_3)^2}{k_2^2k_3^2} + k_{23}^iL^{(2)}_i(\vk_2,\vk_3)\right]. 
\end{align}

Using  the identities 
 $\T D_+^2 = 2 D_+\T D_+ + 2 \dot{D}_+$ and $(\T- \frac{3}{2} H^2)^{-1} [\frac{3}{2}H^2 D_+^3] = \frac{1}{6}D_+^3$, where $D_+$ is the growing solution to $(\T- \frac{3}{2} H^2)D_+=0$, and $H=2/(3t)$, it is straightforward to check that  $\mathcal{A}^{(3)}=\mathcal{B}^{(3)}=
 \mathcal{C}^{(3)}=\mathcal{D}^{(3)}=\mathcal{E}^{(3)}=1$ for EdS.

\end{section}

 \bibliographystyle{JHEP}  
 \bibliography{biblio_FP_MG.bib}

\end{document}